\documentclass[a4paper,11pt]{article}
\usepackage{jcappub}
\usepackage{bm}
\usepackage[utf8]{inputenc}
\usepackage{amsmath}
\usepackage{amssymb}
\usepackage{subeqnarray}
\usepackage{graphicx}
\usepackage{xcolor}
\usepackage{siunitx}
\usepackage{subcaption}
\usepackage[normalem]{ulem}
\usepackage{ragged2e}
\usepackage{aas_macros}

\usepackage{epsfig}

\newcommand\spart{\;\raise1.0pt\hbox{/}\hskip-6pt\partial}
\newcommand\spartb{\;\overline{\raise1.0pt\hbox{/}\hskip-6pt\partial}}

\newcommand{\be}{\begin{equation}}
	\newcommand{\ee}{\end{equation}}
\newcommand{\bea}{\begin{eqnarray}}
	\newcommand{\eea}{\end{eqnarray}}
\newcommand{\beal}{\begin{align}}
	\newcommand{\eeal}{\end{align}}
\newcommand{\beas}{\begin{subeqnarray}}
	\newcommand{\eeas}{\end{subeqnarray}}
\newcommand{\dd}{{\rm d}}
\newcommand{\ii}{{\rm i}}
\newcommand{\HH}{{\cal H}}
\newcommand{\gr}[1]{\boldsymbol{#1}}

\newcommand{\VM}{\Phi}

\abstract{We revisit the presence of primordial gravitational vector modes ($\mathcal{V}$-modes) and their sourcing of primordial magnetic fields (PMF), i.e. magnetogenesis.  As the adiabatic vector mode generically decays with expansion, we consider exotic initial conditions which circumvent this issue and lead to observational imprints. The first initial condition is an isocurvature mode between photons and neutrinos vorticities, and the second one is a non-trivial initial condition on the neutrino octupole. Both types of conditions sustain a constant vector mode on super Hubble scales at early times. We also consider a third scenario in which the adiabatic vector modes are rapidly sourced, at a given early but finite time, by an exotic component which develops an anisotropic stress. We find the best fitting parameters in these three cases to CMB and BAO data. We compare the resulting $B$-mode spectra of the CMB to data from BICEP/Keck and SPTpol. We find that none of the proposed initial conditions can produce large enough PMFs to seed every type of magnetic fields observed. However, $\mathcal{V}$-modes are still consistent with the data and ought to be constrained for a better understanding of the primordial Universe before its hot big-bang phase.}
\begin{document}
	
	\title{Resurrecting Gravitational Vector Modes and their Magnetogenesis}

	\author[a]{Ali Rida Khalife,}
	\emailAdd{ridakhal@iap.fr}
	
	\author[a]{Cyril Pitrou}
	\emailAdd{pitrou@iap.fr}
	\affiliation[a]{Institut d'Astrophysique de Paris, CNRS UMR 7095, 98 bis Bd Arago, 75014 Paris, France.}
\date{\today}
\maketitle
\section{Introduction}
\label{Sec:Intro}
Explaining the origin of astrophysical magnetic fields (MFs) has been a challenge for many years. In galaxies~\cite{MF_Review2,MF_MilkyWay} and galactic clusters~\cite{MF_Review1,MF_Clusters3}, as well as in inter-galactic~\cite{MF_Intergalactic1,MF_Intergalactic2} and inter-cluster~\cite{MF_Intercluster3,MF_Intracluster} media, these MFs have been observed with magnitudes ranging from a few $\mu$G to $\sim10^{-10}\ \mu$G (see~\cite{Review1,Review2} and references therein for more details). One can explain these large magnitudes with astrophysical mechanisms, such as the dynamo mechanism~\cite{Dynamo_DOLGINOV,Dynamo_Kronberg}. Starting from an initial seed field, and in the presence of an electrically conducting medium, the dynamo mechanism describes the transformation of the medium's kinetic energy into MF energy through the act of turbulent motion. 

However, there are two main issues with this mechanism. First, an explanation for the origin of this seed MF and its strength is still needed. Second, for such a dynamo mechanism to work, it needs to be very efficient in amplifying a tiny seed field. This could be possible for a system such as the Milky Way where, even if we start with an extremely small seed field ( $B_\mathrm{seed}\sim10^{-30}$G), assuming maximum efficiency of the dynamo mechanism can explain the $\sim\mu$G field observed.  However, this cannot explain the case for MFs found in high-redshift galaxies~\cite{MF_HighREdshift1,MF_HighRedshift2,MF_HighRedshift3,MF_HighRedshift4}. The latter did not have enough time to go through as many revolutions as low-redshift galaxies, and so $B_\mathrm{seed}$ cannot be amplified to reach the $\mu$G level observed in them. If these two types of galactic MFs have the same origin, then $B_\mathrm{seed}$ should be $\sim10^{-14}$ G, making it slightly difficult to justify.

In order to solve these two problems, one would need a magnetogenesis mechanism that explains the origin of a seed MF with magnitude $B_\mathrm{seed}\sim 10^{-14}$ G. One proposition at astrophysical epochs is called the Biermann-battery mechanism\footnote{A similar mechanism was proposed in~\cite{Turbulant_Dynamo}.}~\cite{Biermann,Dynamo_DOLGINOV,Review1}. Explained briefly, this mechanism occurs in a plasma when its non-uniform rotation creates a misalignment between electron density and pressure (or temperature) gradients (see section 4 of~\cite{Dynamo_DOLGINOV}). This misalignment acts as a source for the time evolution of MFs, thus creating their seed. In its original form, the Biermann-battery mechanism applies to very small (comoving) scales (1-10 kpc), and the resultant $B_\mathrm{seed}\sim 10^{-25}-10^{-24}$ G, making it marginally suitable to explain galactic scale MFs~\cite{Dynamo_Kronberg,Review1}. More recently, the authors of~\cite{Biermann_Yacine} made substantial improvements on previous works~\cite{Biermann_Flitter,Biermann_Naoz} and showed that, under certain initial conditions, this mechanism could be extended to galactic clusters scales (0.1-1 Mpc). However, it remains to be shown whether this mechanism could be further extended to larger scales as observed by electromagnetic cascades of blazars (see section 2 of~\cite{Review2}).\footnote{After finishing the work presented here, it was shown later in~\cite{Garg:2025mcc} the possibility of generating MFs in voids using classical electromagnetism.}

An alternative explanation for the origin of $B_\mathrm{seed}$ is to consider its production in the pre-recombination Universe, i.e. a Primordial Magnetic Field (PMF). Being of early-Universe nature allows it to act as a seed field for more systems than in the case of an astrophysical seed. This subject has been an active field of research for many years~\cite{Hoyle1958,Thorne:1967zz,Durrer:2013pga,Review1,Review2} and was studied in different contexts. The latter includes (but certainly not limited to) PMFs from phase transitions in the early Universe (either the Quantum Chromodynamic or the Electro-Weak one)~\cite{Caprini:2009yp,PMF_EXPT,PMF_PhaseTrans1,PMF_PhaseTran2,PMF_PhaseTran3,PMF_QCD,PMF_QCD2,PMF_QCD3,PMF_QCD4}, including the production of magnetic monopoles~\cite{PMF_Monopole1,PMF_Monopole2,PMF_Monopole3}; PMFs from inflation~\cite{PMF_Inflation1,PMF_Inflation2,PMF_Inflation3,PMF_Inflation4,PMF_Inflation5}, with the production of a helical~\cite{PMF_Helicity,PMF_Helicit2,PMF_Helicity4,PMF_Helicit3} component and in connection to baryogenesis~\cite{PMF_Baryogen1,PMF_Baryogen2,PMF_Baryogen3}; PMFs from cosmic defects~\cite{PMF_Defects1,PMF_Defects2,PMF_Defects3,PMF_Defects4} and PMFs from second order perturbations~\cite{PMF_2ndOrder1,PMF_2ndOrder2,PMF_2ndOrder3,PMF_2ndOrder4,PMF_2ndOrder5,PMF_2ndOrder6}.  In addition, a great amount of Magnetohydrodynamic (MHD) simulations have been conducted in the presence of PMFs while being agnostic about the exact magnetogenesis mechanism~\cite{MHD1,MHD2,MHD3,MHD4,MHD5,MHD6} (see section 4 of~\cite{Review2} and references therein for more details). All these different mechanisms have rich phenomenology, and produce signatures in one of the best cosmological probes, the Cosmic Microwave Background (CMB). Moreover, PMFs have been shown to be promising in reducing the Hubble Tension, by increasing the clumpiness of the plasma on small scales which then leads to an earlier recombination and a shorter sound horizon. This property makes it another motivation to explore PMFs in more detail (see~\cite{PMF_H0} and references therein for more details). 

There are two direct ways in which MFs impact the CMB. The first is through Faraday rotation, where MFs rotate the CMB polarization, converting part of the $E$-modes into $B$-modes~\cite{Faraday1,Faraday2,Faraday3,Faraday4}. The second is via the anisotropic stress of the MF, which depends quadratically on it and whose characteristics also depend on the magnetogenesis mechanism, as it influences the evolution of metric perturbations. In particular, scalar, vector and tensor perturbations, which themselves have their own signatures on the CMB~\cite{ZaldarriagaAndSeljak,Lewis:2004kg,KamionkowskiAndKosowsky,HuAndWhite}, interact differently with the anisotropic stress of the MF. For instance, a non-vanishing helicity for the MF produces different signatures on the CMB spectra depending on the type of perturbations (see~\cite{Helicity_CMB1,Helicity_CMB2,Helicity_CMB3,Helicity_CMB4,Planck:2015} and references therein for more details). This topic has also been studied theoretically~\cite{MF_Lewis,Kosowsky_etal:2002,Paoletti_theory,Caprini:2003vc,Finelli_Scalars2008,Ichiki_etal:2008gr,Takahashi:2007ds,Durrer:1998ya,Kahniashvili:2005xe}, and later constraints on the MF's power-spectrum from CMB data also presented~\cite{Paoletti:2022gsn,Galli:2021mxk,Paoletti_Forecasts,Itchiki_etal2006bq,Minoda:2020bod}. Note that all these mentioned works assume the presence of a stochastic background of PMFs with the three types of perturbations, without specifying a magnetogenesis mechanism from the latter. On the other hand, some works did consider magnetogenesis from non-trivial second-order primordial perturbations~\cite{Matarrese:2004kq,PMF_2ndOrder5,PMF_2ndOrder3}, from cosmic strings defects~\cite{Itchiki_2ndOrder,Itchiki_CosmicStrings_Vmodes} or from an isocurvature initial condition (IC) between photons and neutrinos for vector perturbations~\cite{Itchiki_V-modes} (see also~\cite{Lewis:2004kg} for the impact this type of ICs has on the CMB).

Building on the above findings, this work presents new insights on magnetogenesis from non-trivial ICs of gravitational vector modes ($\mathcal{V}$-modes)\footnote{Not to be confused with the Stokes parameter V, often named V mode, describing circular polarization.}. The main mechanism considered is based on the presence of vorticity in the decoupled plasma, which induces an electric field and subsequently sources a magnetic field due the Maxwell-Faraday structure relation. This vorticity can be generated either from $\mathcal{V}$-modes at linear order in perturbation theory, or from scalar perturbations at second order. However, as was shown in~\cite{Matarrese:2004kq,PMF_2ndOrder5,PMF_2ndOrder3}, second order perturbations are too small to generate enough vorticity to source PMFs. Therefore, we focus on the generation of PMFs from $\mathcal{V}$-modes at linear order in perturbation theory. In this mechanism, PMFs are produced solely from $\mathcal{V}$-modes and we do not consider any additional magnetogenesis mechanisms. Hence, the $\mathcal{V}$-modes will both leave a trace on the CMB and produce PMFs, but the latter's impact on the CMB is negligible (the reason for this will be evident later), that is we can ignore the direct influence of the PMFs on the CMB.

We start by a brief theoretical overview of $\mathcal{V}$-modes and MFs in curved spacetime in Section~\ref{Sec:Theory}. Then, in addition to the isocurvature IC between neutrinos and photons~\cite{Itchiki_V-modes,Lewis:2004kg,Rebhan:1991sr,Rebhan:1994zw}, we consider non-zero IC on the neutrino octupole and a third IC we call the Sourced Mode (SMD). In the latter, an exotic species' anisotropic stress directly sources the $\mathcal{V}$-modes suddenly at a given early but finite time of the primordial Universe.  We remain agnostic on the origin of these ICs, and we treat them in a phenomenological way, as elaborated in Section~\ref{Sec:ICs}. Moreover, we find the best-fit parameters for each IC to  Planck 2018 TT, TE, EE +lowE+lensing~\cite{Planck_Primary,Planck_Lensing}, SPT-3G 2018 TT, TE and EE~\cite{Dutcher_etal,Balkenhol_etal} and Baryon Acoustic Oscillations (BAO) data from the 6 degree Field Galaxy Survey (6dFGS)~\cite{6dFGS}, BOSS~\cite{BOSS_DR7,BOSS_DR12} and eBOSS~\cite{eBOSS_DR16} surveys (which we collectively label BAO). To be specific, we constrain the parameters associated to $\mathcal{V}$-modes with these data, and derive the shape of the PMF power spectrum from them. We also study the co-existence of both tensor and $\mathcal{V}$-modes in each IC case. We present the results, including the produced PMF from each IC, in the corresponding subsections of Section~\ref{Sec:ICs}. We present a brief comparison of the CMB power spectra between the different ICs in Section~\ref{Sec:CMB_Spectra} before concluding in Section~\ref{Sec:Conclusion}.

It is worth emphasizing that we are aware of the challenges these ICs face with our current understanding of the early Universe. In particular, the isocurvature and octupole ICs are difficult to motivate given that before the MeV scale neutrinos are tightly self-coupled and also tightly coupled to the electron-positron plasma. Moreover, although the presence of Dark species in the early Universe is predicted by theories such as String Theory~\cite{String1,String2,String3,String4} and Supersymmetry~\cite{SUSY1,SUSY2,SUSY3,SUSY4}, yet for these species to have an abrupt rise in their anisotropic stress is more difficult to explain. However, we are considering such ICs precisely to highlight the difficulty of generating PMFs from $\mathcal{V}$-modes, and to motivate further theoretical investigation of these ICs.

As a summary of our results, we can say that for each IC considered here (either with or without tensor modes), the resultant best-fit magnetic fields have a root-mean-square value (smoothed with a Gaussian window function over 1Mpc) $\sim10^{-29}-10^{-26}$G. These values are too small to explain all astrophysical MFs observed, making magnetogenesis from these ICs less likely. Even if we fix the $\mathcal{V}$-modes' amplitude to extremely large values, one still needs another magnetogenesis mechanism. However, the resultant $\mathcal{V}$-modes from these ICs have an appreciable impact on the CMB,  with some of them being still consistent with the $B$-modes measured by BICEP/Keck and SPTpol (see section~\ref{Sec:ICs} for more details). Although a further theoretical explanation for the presence of these ICs is needed, this result is an indication that we should not ignore the effect of $\mathcal{V}$-modes when drawing cosmological conclusions from CMB $B$-mode measurements. 

To perform our computations, we modified the Boltzmann code \texttt{CLASS}~\cite{CLASSI,CLASSII}\footnote{A public release of this version will be available upon publication of the article.} to incorporate $\mathcal{V}$-modes and each of the above mentioned ICs. The Boltzmann hierarchy for vector modes is implemented with the total angular momentum approach~\cite{HuAndWhite,TAM2}, reviewed hereafter, following the implementation for tensors used in~\cite{Pitrou:2020lhu}.  Moreover, to find the best-fitting parameters, we use the \texttt{BOBYQA} algorithm~\cite{BOBYQA_I,BOBYQA_II} through its implementation in the Bayesian analysis code \texttt{COBAYA}~\cite{COBAYA1,COBAYA2,COBAYA3}. 

\section{Theory}
\label{Sec:Theory}

In this section, we elaborate briefly on the theory behind $\mathcal{V}$-modes and how they produce PMFs. The following is based on the Refs.~\cite{HuAndWhite,PitrouModes,Peter-Uzan,PMF_2ndOrder3}, to which we direct the interested reader for more details. 

\subsection{Perturbations}
Let the metric of spacetime (with flat spatial slices) be a perturbed Friedmann-Lema\^itre-Robertson-Walker (FLRW) metric:
\be 
g_{\mu\nu} = a^2(\eta)\big(\eta_{\mu\nu}+h_{\mu\nu}\big),
\label{Eq:Metric}
\ee
where $\eta_{\mu\nu}=\mathrm{diag}[-1,1,1,1]$ is the Minkowski metric of flat spacetime, $h_{\mu\nu}$ its perturbation and $a(\eta)$ the scale factor as a function of conformal time, $\eta$. Since we are interested in $\mathcal{V}$-modes, we will focus on the latter's perturbations and ignore scalar and tensor ones, for brevity. We choose the gauge in which the vector perturbation is in the $(0-i)$ component of the metric, so that
\be
h_{00}=0; \ h_{0i} = -\Phi_i; \ h_{ij}=0, 
\label{Eq:Pert_Metric}
\ee
where $\Phi_i$ is the (divergenceless) V-mode perturbation satisfying $D^i\Phi_i=0$, with $D_i$ being the covariant derivative of the flat spatial background metric.

The spacetime evolution of $\Phi_i$ is given by the $(0-i)$ and $(i-j)$ components of the Einstein equations for vector modes~(see eqs.~(5.113-5.114) of~\cite{Peter-Uzan}). Written explicitly,\footnote{In units where the reduced Planck constant $\hbar$ and the speed of light $c$ are unity.} the latter is given by
\be
D_{(i}{\Phi}_{j)}'+2\mathcal{H}D_{(i}\Phi_{j)} = 8\pi Ga^2 \sum_{s} \bar{p}_s (\pi_s)_{ij}
\label{Eq:0-iComponent}
\ee
where parentheses around indices denote symetrization, $'$ denotes a derivative with respect to (w.r.t) $\eta$, $\mathcal{H}={a}'/a$, $G$ is Newton's constant, $\bar{p}_s$ is the pressure of species $s$ (at the background level) and $(\pi_s)_{ij}$ is its dimensionless anisotropic stress. We also used the definition of the dimensionless anisotropic stress of a given species as
\be\label{Defpiij}
T^s_{ij} \equiv \bar{p}_s (\pi_s)_{ij}.
\ee
On the other hand, the $(0-i)$ component of the Einstein equations is 
\be
\Delta\Phi_i = -16\pi Ga^2\sum_{s}(\bar{\rho}_s+\bar{p}_s)(\tilde{v}_s)_i
\label{Eq:i-jComponent}
\ee  
where $\Delta = D^iD_i$ is the Laplace operator, 
$(\tilde{v}_s)_i = (v_s)_i -\Phi_i$ with $(v_s)^i$ being the divergenceless part of the velocity of species $s$ (hence $D_i (v_s)^i = 0$), and $\bar{\rho}_s$ is the energy density of species $s$ (at the background level). 
Since for a given species the spatial components of the velocity is defined as $(u_s)^i = a^{-1} (v_s)^i$, then $(v_s)^i$ is the velocity seen by observers with constant spatial coordinates. Similarly, since $(u_s)_i = a (\tilde{v}_s)_i$, then $(\tilde{v}_s)_i$ is the spatial velocity seen by an observer whose velocity is normal to constant-$\eta$ hypersurfaces. This is why we use $\tilde{v}_s$ in describing the dynamics in subsequent sections.

\subsection{Normal mode expansion}
\label{Sec:Norm_Modes_Exp}
\subsubsection{Mode functions}
\label{Sec:Mode_Function}
The dynamics of perturbations is conveniently analyzed using a normal mode expansion. Normal modes are eigenfunctions of the Laplace operator, and for vector perturbations they satisfy\footnote{The overbars on normal modes indicate that we consider the propagating direction convention detailed in section~7.1 of~\cite{PitrouModes}.}
\be
\big(\Delta +k^2\big)\bar{Q}^{\pm}_i=0
\label{Eq:Laplace}
\ee
with $D^i \bar{Q}^{\pm}_i=0$. Using a spherical coordinates system, with the associated orthonormal basis, $(\gr{e}_r,\gr{e}_\theta,\gr{e}_\phi)$, one can write the wave-vector of a mode $k$ as $\gr{k}=k\gr{e}_r$. Then a basis for the transverse plane can be built from $\gr{e}_1 = \gr{e}_\theta$ and $\gr{e}_2 = \gr{e}_\phi$ (which reduce to $\gr{e}_x$ and $\gr{e}_y$ when the mode is aligned with $\gr{e}_z$ as in~\cite{HuAndWhite}) as
\be
\gr{e}_{\pm} = \frac{1}{\sqrt{2}}\left(\gr{e}_1 \mp \ii \gr{e}_2\right).
\label{Eq:Basis_vec}
\ee
A particularly useful representation of the mode function is\footnote{We use a different sign convention compared to~\cite{HuAndWhite}, as we follow~\cite{PitrouModes}. Spherical harmonics in both references differ by a factor $(-1)^m$, hence the sign difference in~\eqref{Eq:Q^pm} so that~\eqref{Qini} holds in both conventions.}
\be
 \bar{Q}^{\pm}_j(\gr{k},\gr{x}) = \pm \ii  \gr{e}^\mp_j {\rm e}^{\ii \gr{k} \cdot \gr{x}} 
\label{Eq:Q^pm}
\ee
with which we can expand $\Phi_i$ as
\be
\Phi_i(\gr{x}) = \int \frac{\dd^3 \gr{k}}{(2\pi)^3} \sum_{m=\pm1} \Phi^{(m)}(\gr{k})
 \bar{Q}^{(m)}_i(\gr{k},\gr{x}).
 \label{Eq:Phi_expan_Mode}
\ee
The velocities $(v_s)_i$ and $(\tilde{v}_s)_i$, being divergenceless as well, are expanded in the same way, hence defining $v_s^{(m)}$ and $\tilde{v}_s^{(m)}$ for $m = \pm1$.

We can further build normal modes for tensor-valued perturbations of the vector type with
\be
\bar{Q}^{(m)}_{ij} = -\frac{1}{2k} \left(D_i \bar{Q}^{(m)}_j + D_j \bar{Q}^{(m)}_i\right).
\label{Eq:Basis_anisotropic}
\ee
In particular, we can expand it as
\be
\pi_{ij}(\gr{x}) = \int \frac{\dd^3 \gr{k}}{(2\pi)^3} \sum_{m=\pm1} \pi^{(m)}(\gr{k})
 \bar{Q}^{(m)}_{ij}(\gr{k},\gr{x}).
 \label{Eq:Pi_expan_Mode}
\ee

Inserting eqs.~\eqref{Eq:Phi_expan_Mode} and~\eqref{Eq:Pi_expan_Mode} into eq.~\eqref{Eq:0-iComponent}, we find
\be
\Phi^{(m)}{}' + 2{\cal H}\Phi^{(m)} = -\frac{8\pi G a^2}{k}\sum_{s} \bar{p}_s \pi_s^{(m)},
\label{Eq:Evol_Phi_ModeFunction}
\ee
which will be our reference equation for the evolution of $\mathcal{V}$-modes. For completeness, applying the same logic to the constraint eq.~\eqref{Eq:i-jComponent}, we find
\be
\Phi^{(m)} = \frac{16\pi Ga^2}{k^2}\sum_{s}(\bar{\rho}_s+\bar{p}_s) \tilde{v}_s^{(m)}.
\label{Eq:Constraint_Eq}
\ee

\subsubsection{Total angular momentum for photons}

The photon temperature fluctuation $\Theta$ is not a pure tensor, since at each point of spacetime it also depends on the propagating direction $\bar{\gr{n}}$ (which is opposite to the observing direction). Since a position on the sky is given by $-\chi \bar{\gr{n}}$, where $\chi$ is the comoving distance, there will be an additional angular dependence when expanding $\Theta$ in modes, due to the direction of the plane wave with respect to the observer. Hence this implies that the angular dependence  of $\Theta$ is twofold, and they can be combined in a total angular momentum basis~\cite{HuAndWhite,PitrouModes} given by the functions
\be\label{DefGslm}
{}_s \bar{G}^{(\ell' m)}(\gr{k},\chi,\bar{\gr{n}}) = \sum_{\ell} \bar{c}_{\ell} \,{}_s\bar{\alpha}_{\ell}^{(\ell'm)}(k\chi) {}_s Y_{\ell}^m(\bar{\gr{n}})\,.
\ee
The prefactors are defined as
\be
\bar{c}_{\ell} = (-i)^{\ell}\sqrt{4\pi(2\ell+1)}
\ee
and the ${}_s Y_{\ell}^m(\bar{\gr{n}})$ are the spin-weighted spherical harmonics~\cite{Spherical_Harmonics1,Spherical_Harmonics2} with spin $s$, and ${}_s \bar{\alpha}_{\ell}^{(\ell'm)}$ are radial functions given explicitly in eqs.~\eqref{Eq:s_alpha^lm_l'1}-\eqref{Eq:Wigner3j}~\cite{PitrouModes}. Since $\Theta$ transforms as a scalar under general coordinate transformations, we should then set $s=0$.

The relation between the functions~\eqref{DefGslm} and the normal modes (eq.~\eqref{Eq:Q_G_relation}) and convenient contractions by the propagation direction (eqs.~\eqref{Eq:Num_Facs_Q_G}-\eqref{Eq:hat_n_ij_pm}) give the most needed relations (suppressing the normal modes' dependence on $\gr{k}$ and $\gr{x} = - \chi \bar{\gr{n}}$ for brevity)
\bea
\bar{Q}^{(m)}_i \bar{n}^i &=& {}_0 \bar{G}^{(1\,m)}(\gr{k},\chi,\bar{\gr{n}})\label{Qini}\\
\bar{Q}^{(m)}_{ij} \bar{n}^i \bar{n}^j &=& \frac{1}{\sqrt{3}}\,{}_0 \bar{G}^{(2\,m)}(\gr{k},\chi,\bar{\gr{n}})\label{Qijnij}.
\eea
Using these definitions, the photon temperature is therefore expanded as
\be\label{ExpTheta}
\Theta =  \int
\frac{\dd^3 \gr{k}}{(2\pi)^3} \sum_{\ell} \sum_{m=\pm 1}
\Theta_{\ell}^{(m)}(\gr{k})
{}_0 \bar{G}^{\ell m}(\gr{k},\chi,\bar{\gr{n}})\,.
\ee
On the other hand, the polarization Stokes parameters $Q$ and $U$, which are components of the spin-2 polarization tensor, can be expanded as 
\bea
&&Q \pm \ii U =  \\
&& \int
\frac{\dd^3 \gr{k}}{(2\pi)^3} \sum_{\ell} \sum_{m=\pm 1}
(E_{\ell}^{(m)} \pm \ii B_{\ell}^{(m)})
{}_{\pm 2} \bar{G}^{\ell m}\nonumber
\eea
given in terms of the famous $E$ and $B$ modes of the CMB. 

Finally since the anisotropic stress of all species enters the Einstein equation eq.~\eqref{Eq:Evol_Phi_ModeFunction}, we need to relate the photon's anisotropic stress, $\pi_\gamma^{(m)}$, to $\Theta_\ell^{(m)}$. The energy density fluctuations of photons propagating in direction $\bar{\gr{n}}$ is $4 \bar{\rho}_\gamma \Theta(\bar{\gr{n}})$, hence (see e.g. section 1 of \cite{Pitrou:2008hy})
\be
T^\gamma_{ij} = 4 \bar{\rho}_\gamma \int \frac{\dd^2 \bar{\gr{n}}}{4\pi} \Theta(\bar{\gr{n}})\left(\bar{n}^i \bar{n}^j - \frac{1}{3}\delta^{ij}\right)
\ee
where $\delta_{ij}$ is the Kronecker delta. Using the orthogonality relation eq.~\eqref{Eq:ortonormality_n_ij}, we see that when inserting the property~\eqref{Qijnij} into the temperature's expansion~\eqref{ExpTheta}, the directional integration selects only the quadrupolar dependence ($\ell=2$). Given the definition~\eqref{Defpiij} and its expansion~\eqref{Eq:Pi_expan_Mode} we then get
\be\label{Eq:Pim_def}
\pi_\gamma^{(m)} = \frac{8}{5}\sqrt{3} \Theta_2^{(m)}\,.
\ee
A similar method based on the evaluation of $T_{0i}$ allows to show that the dipole matches exactly the velocity, that is
\be\label{relationsvTheta1}
\Theta_1^{(m)} = v_\gamma^{(m)}\,,
\ee
such that
\be\label{relationsvTheta2}
\tilde{v}_\gamma^{(m)} = \Theta_1^{(m)} - \Phi^{(m)}\,,
\ee
which enters the constraint equation~\eqref{Eq:Constraint_Eq}.

\subsubsection{Boltzmann hierarchy for photons}

The time evolution of the temperature multipoles is given by the Boltzmann hierarchy~\cite{HuAndWhite,PitrouModes}:

	\bea\label{Eq:Hierarchy}
	(\Theta_{\ell}^{(m)}) {}' &=& k\left[\frac{{}_0 \kappa_{\ell}^m}{2{\ell}-1}
	\Theta_{{\ell}-1}^{(m)} - \frac{{}_0 \kappa_{{\ell}+1}^m}{2{\ell}+3}
	\Theta_{{\ell}+1}^{(m)} \right] +{}^\Theta {\cal C}_{\ell}^m -\tau' \Theta_{\ell}^{(m)} \,, \ (\ell\geq m)\,,\\
	(E_{\ell}^{(m)}){}' &=& k\left[\frac{{}_2 \kappa_{\ell}^m}{2{\ell}-1}
	E_{{\ell}-1}^{(m)} - \frac{{}_2 \kappa_{{\ell}+1}^m}{2{\ell}+3}
	E_{{\ell}+1}^{(m)} -\frac{2m}{{\ell}({\ell}+1)} 
	B_{\ell}^{(m)}\right] +{}^E {\cal  C}_{\ell}^m -\tau' E_{\ell}^{(m)} \,,\ (\ell\geq2)\,,\nonumber\\
	(B_{\ell}^{(m)}){}' &=& k\left[\frac{{}_2 \kappa_{\ell}^m}{2{\ell}-1}
	B_{{\ell}-1}^{(m)} - \frac{{}_2 \kappa_{{\ell}+1}^m}{2{\ell}+3}
	B_{{\ell}+1}^{(m)} +\frac{2m}{{\ell}({\ell}+1)}  E_{\ell}^{(m)}\right] -\tau' B_{\ell}^{(m)}\,, \ (\ell\geq2). \nonumber
	\eea
 A few definitions are in order. First,
 \be\label{Defkappaslm}
 {}_s \kappa_\ell^m \equiv\sqrt{\frac{(\ell^2-m^2)(\ell^2-s^2)}{\ell^2}},
 \ee
 for $\ell\geq1$ (it equals 0 for $\ell=0$) and $\tau'\equiv an_e\sigma_\mathrm{T}$ is the time derivative of the optical depth, $\tau$, $n_e$ is the number density of (free) electrons and $\sigma_\mathrm{T}$ is the Thomson cross section. Second, the source terms ${}^X{\cal C}^m_{\ell}$, for $X=\{\Theta,E\}$, account for gravitational effects on the CMB, in addition to the anisotropic nature of Compton scattering. These are given by
 \begin{align}
 {}^{\Theta}{\cal C}_{\ell}^m &= \left(\tau' v_b^{(m)}+(\Phi^{(m)}){}'\right)\delta_{\ell1}+\tau'P^{(m)}\delta_{\ell2},
 \label{Eq:Source_Theta}\\
 {}^{E}{\cal C}_{\ell}^m &= -\sqrt{6}P^{(m)}\delta_{\ell2},
 \label{Eq:Source_E}
 \end{align}
 where
 \be
 P^{(m)} \equiv \frac{1}{10}\left(\Theta_2^{(m)} - \sqrt{6}E_2^{(m)}\right).
 \label{Eq:P^m}
 \ee

\subsubsection{System of baryons and photons}

The general Euler equation for vector modes, when considering an imperfect fluid which does not interact with other species, is\footnote{Note that in deriving this relation, one needs to use the continuity equation as well.}
\be
 \left[a^4(\bar \rho_f +\bar p_f) \tilde v_f^{(m)}\right]' = -\frac{a^4
 	\bar{p}_f}{2} k \pi_f^{(m)}.
 \label{Eq:GenEuler}
 \ee 
It is then checked using eqs.~\eqref{Eq:Pim_def}, \eqref{relationsvTheta2} and $\bar{p}_\gamma = \bar{\rho}_\gamma/3 \propto 1/a^4$ that the first multipole ($\ell=1$) in the $\Theta$ hierarchy of eq.~\eqref{Eq:Hierarchy} corresponds to the evolution of the photon velocity, if supplemented with the collision term with baryons. It then simply reads
\be
 \tilde{v}_{\gamma}^{(m)}{}' = -k\frac{\sqrt{3}}{5}\Theta_2^{(m)}+\tau{}'\delta v^{(m)}
 \label{Eq:Evol_Photon_Dipole}
 \ee
where the baryon-photon velocity difference is defined as
 \be
\delta v^{(m)} \equiv {\tilde v}^{(m)}_b - {\tilde v_{\gamma}}^{(m)}= v_b^{(m)} - v_{\gamma}^{(m)}.
\label{Eq:V}
\ee

From the conservation of the total stress-energy tensor, we can deduce the effect of Compton scattering on baryons. Using the fact that the latter are well described by a perfect ($(\pi_b)_{ij}=0$) and pressureless ($\bar{\rho}_b \propto 1/a^3$) fluid, the baryon Euler equation takes the simple form
\be
(\tilde{v}_b^{(m)}){}' + \HH \tilde{v}_b^{(m)} = -\frac{\tau{}'}{R}\delta v^{(m)},
\label{Eq:Evol_baryon}
\ee
where 
\be
R\equiv \frac{3\bar{\rho}_b}{4\bar{\rho}_{\gamma}}\,.
\ee

As a side note, since before recombination baryons and photons are tightly coupled, they can be described collectively as an imperfect fluid whose velocity is
\be
\tilde v_f^{(m)} \equiv \frac{R}{1+R} \tilde v_b^{(m)} + \frac{1}{1+R} \tilde v_\gamma^{(m)}.
\label{Eq:Def_vf}
\ee
By combining the Euler eqs.~\eqref{Eq:Evol_Photon_Dipole} and \eqref{Eq:Evol_baryon} with the weights eq.~\eqref{Eq:Def_vf}, we get the Euler equation of this imperfect fluid description
\be
[(1+R)\tilde{v}_f^{(m)}]' = -\frac{k}{8} \pi_\gamma^{(m)} = -\frac{\sqrt{3} k}{5} \Theta_2^{(m)}\,,
\label{Eq:Euler_Photon}
\ee
which is equivalent to the general Euler equation eq.~\eqref{Eq:GenEuler} with $\bar{\rho}_f = \bar{\rho}_b + \bar{\rho}_\gamma$, $\bar{p}_f = \bar{p}_\gamma$ and $\pi^{(m)}_f = \pi^{(m)}_\gamma$. 

The equation satisfied by the velocity difference, eq.~\eqref{Eq:V}, is then found from the difference of the photon and baryon Euler equations. It can be written in a form suitable for a tight-coupling approximation (TCA)~\cite{Pitrou:2010ai} 
\be
\begin{split}
-\frac{\tau' \delta v^{(m)}}{R} &= \frac{\delta v^{(m)}{}' + \HH\left[\tilde v_f^{(m)} +\frac{\delta v^{(m)}}{(1+R)}\right]}{1+R}\\
& -\frac{\sqrt{3} k \Theta_2^{(m)}}{5(1+R)}
\end{split}
\label{Eq:TCV}
\ee
which we will frequently refer to later. 

 \subsubsection{Neutrino hierarchy}

The neutrinos, being relativistic like the photons, are described by a set of multipoles ${\cal N}_\ell^{(m)}$.  The main difference is that, after they decouple, neutrinos travel freely, and thus they will not have any interactions terms. Therefore, the evolution hierarchy can be read off the first equality in eq.~\eqref{Eq:Hierarchy} by removing the terms proportional to $\tau{}'$. The result is
\be
({\cal N}_{\ell}^{(m)}){}' = k\left[\frac{{}_0 \kappa_{\ell}^m}{2{\ell}-1}{\cal N}_{{\ell}-1}^{(m)} - \frac{{}_0 \kappa_{{\ell}+1}^m}{2{\ell}+3}{\cal N}_{{\ell}+1}^{(m)} \right] + (\Phi^{(m)}){}'\delta_{\ell1}, \ (\ell\geq m).
\label{Eq:Nu_Hierarchy}
\ee

Moreover, the neutrino anisotropic stress, which enters the Einstein equation~\eqref{Eq:Evol_Phi_ModeFunction}, is related to ${\cal N}_2^{(m)}$ via a relation analogous to eq.~\eqref{Eq:Pim_def}, while the velocity $\tilde{v}_\nu$ which enters the constraint equation~\eqref{Eq:Constraint_Eq} is found via a relation analogous to eq.~\eqref{relationsvTheta2}.

The equations presented in this subsection provide a system of differential equations describing the impact $\mathcal{V}$-modes have on the CMB. We can now move to the production of PMFs from $\mathcal{V}$-modes. As mentioned in the Introduction, the presence of PMFs creates an additional source of anisotropic stress that will back-react on the $\mathcal{V}$-modes and the CMB hierarchy. However, we are not considering such back-reaction since it is negligible, as we will see in Section~\ref{Sec:ICs}. Therefore, we first solve the $\mathcal{V}$-modes' evolution, and then use that as a source for PMFs, as we now detail.

\subsection{Magnetic Field Production}
\label{Sec:PMF_Prod} 

The generation of the magnetic field proceeds in two steps. First, as we shall review briefly, Compton collisions between electrons and photons in the plasma create an electric field that forces electrons and protons to behave collectively as a single fluid (the baryons). Then, from the Maxwell-Faraday equations expressed in curved spacetime, this electric field sources a magnetic field if the velocities generating the former are not curl free. The details of the electric field generation, and the subsequent magnetic field generation, is studied in~\cite{PMF_2ndOrder3} (and references therein) up to second order in cosmological perturbations. Here, we will simplify this formalism to first order in perturbations, and will briefly describe the leading steps. 

\subsubsection{Electric field production}

Let us write the Euler equation for the electrons and protons individually, taking into account the presence of Compton interactions with photons and the electric force. They read respectively
\be
\frac{\bar{\rho}_e}{a}[(\tilde{v}_e^i)' + \HH \tilde{v}_e^i] = \bar{n}_e \sigma_T \frac{4 \bar{\rho}_\gamma}{3} (\tilde{v}_\gamma^i -\tilde{v}_e^i) - e \bar{n}_e {\cal E}^i\,,
\label{Eq:Euler_e}
\ee
\be
\frac{\bar{\rho}_p}{a}[(\tilde{v}_p^i)' + \HH \tilde{v}_p^i] = \bar{n}_p \sigma_T \frac{m_e^2}{m_p^2}\frac{4 \bar{\rho}_\gamma}{3} (\tilde{v}_\gamma^i -\tilde{v}_p^i) + e \bar{n}_p {\cal E}^i\,,
\label{Eq:Euler_p}
\ee
where ${\cal E}^i$ is the $i^{\text{th}}$ component of the electric field. The cross section of Compton scattering on protons is reduced by a factor $(m_e/m_p)^2$ compared to the one on electrons, which allows us to neglect this term hereafter since $m_e\ll m_p$. Furthermore, the charge separation of electrons and protons shall remain very small, allowing us to approximate $n_p \simeq n_e$. In addition, it was shown by order of magnitude estimates in~\cite{PMF_2ndOrder3} that the velocities of protons and electrons remain very close due to the generated electric field. Hence, we can define an average baryon velocity by
\be
\tilde{v}_b^i \equiv \frac{\bar{\rho}_e \tilde{v}_e^i+ \bar{\rho}_p \tilde{v}_p^i}{\bar{\rho}_e + \bar{\rho}_p}\,.
\ee
With this definition, and adding the electron and proton Euler equations, we recover the Euler equation for baryons eq.~\eqref{Eq:Evol_baryon} under the approximation $\bar{\rho}_b \simeq \bar{\rho}_p$. On the other hand, taking the difference between eqs.~\eqref{Eq:Euler_e} and~\eqref{Eq:Euler_p} leads to 
\bea
&&(\tilde{v}_{ep}^i)'+\HH \tilde{v}_{ep}^i=\\
&&\frac{4 \bar{\rho}_\gamma}{3 \bar{\rho}_e} a \bar{n}_e \sigma_T (\tilde{v}_\gamma^i - \tilde{v}_e^i) - ea {\cal E}^i \bar{n}_e \left(\frac{1}{\bar{\rho}_e}+\frac{1}{\bar{\rho}_p}\right)\nonumber
\eea
where $\tilde{v}_{ep}^i = \tilde{v}^i_e-\tilde{v}_p^i$. As this quantity remains very small, we can neglect the left hand side of the previous equation, and approximate $\tilde{v}_e^i\simeq\tilde{v}_b^i$, Therefore, using the approximation $\bar{\rho}_e \ll \bar{\rho}_p$, the electric field can be simply expressed as
\be\label{Eq:Ei}
{\cal E}^i \simeq \frac{4 \bar{\rho}_\gamma \sigma_T}{3 e}\left(\tilde{v}_\gamma^i - \tilde{v}_b^i\right).
\ee
This expression corresponds to a state of quasistatic equilibrium. The imbalance of Compton scattering between electrons and protons, which is mainly efficient on the former and tends to separate them from protons, is compensated by an electric force which drives the electrons and protons toward each other until they have the same velocity. In other words, the photons push on the electrons, but this interaction is also felt by the protons which are bound to track the electrons via the electric field which is sustained by the tiny charge separation. Hence, even though the Compton drag is nearly only on the electrons, the inertia is essentially due to the much heavier protons which are bound to follow the electrons via this electric interaction. In Ref.~\cite{PMF_2ndOrder3} it was shown that the timescale to reach this equilibrium is much smaller than cosmological timescales, hence it is justified to assume that the equilibrium $\tilde{v}_{ep}^i\simeq 0$ is preserved at all times. Note that this mechanism does not depend on the fraction of free electrons, hence it is efficient even after recombination. However, its efficiency decays as photons are diluted and their drag decreases.

Finally, just like the velocities, the electric field can be decomposed on normal modes as in eq.~\eqref{Eq:Phi_expan_Mode}. The quasistatic equilibrium relation then takes the form
\be
{\cal E}^{(m)} = \frac{4 \bar{\rho}_\gamma \sigma_T}{3 e}\left(\tilde{v}_\gamma^{(m)} - \tilde{v}_b^{(m)}\right)\,,
\ee
or, using the definition $x_e \equiv \bar{n}_e/(\bar{\rho}_b/m_p)$ for the fraction of free electrons, this can be rewritten in a form suitable for eq.~\eqref{Eq:TCV},
\be
{\cal E}^{(m)}=- \frac{m_p}{ae x_e}\frac{\tau'\delta v^{(m)}}{R},
\label{Eq:E_useful}
\ee
which we will use in later sections.

\subsubsection{Magnetic field sourcing}

Maxwell's structure equations, when written in terms of the Faraday tensor $F_{\mu\nu}$, take the form
\be
\nabla_{[\lambda}F_{\mu\nu]}=0\,.\label{Eq:Fuv_eqs}
\ee
Since there are no electric and magnetic fields at the background level, $\nabla_{\lambda}$ above needs only to be evaluated on the background FLRW geometry, thus keeping all expressions up to linear order. Maxwell-Faraday equation then takes the simple form
\be
(a^2 {\cal B}^i)' = - a^2 \epsilon^{ijk} D_j {\cal E}_k\,,
\ee
where ${\cal B}^i$ is the $i^{\text{th}}$ component of the magnetic field and $\epsilon^{ijk}$ is the Levi-Civita symbol.

Expanding the magnetic field on normal modes as in eq.~\eqref{Eq:Phi_expan_Mode}, and using $\epsilon^{ijk} D_j \bar{Q}^{(m)}_k = \pm k \bar{Q}^{(m)}_i$, we find
\be
(a^2 {\cal B}^{(m)})' = -m k a^2{\cal E}^{(m)}.
\label{Eq:B_final}
\ee
When there is no electric field, the magnetic field decays adiabatically as $1/a^2$. However, when there is a non vanishing electric field initially, the magnetic field is sourced and can be sustained for longer periods of time before it decays. The task then becomes to find suitable ICs in the primordial Universe which lead to this physical situation, which is our next topic.

\section{Initial Conditions}
\label{Sec:ICs}

In this section, we consider three different initial conditions, the first of which has been considered in previous works~\cite{Lewis:2004kg,Itchiki_V-modes,Paoletti_theory,Rebhan:1991sr,Rebhan:1994zw}. As mentioned in the Introduction~\ref{Sec:Intro}, a fair disclaimer must be made about these ICs as they require unconventional conditions. The three ICs presented below are based on the implicit assumption that the fundamental theory of particle interactions goes beyond the Standard Model (SM) of particle physics. Specifically for the isocurvature IC discussed first, one can imagine new interactions that alter the weak nuclear force in a way that decouples neutrinos with a velocity different from that of photons. Alternatively, it could be possible to have interactions that renders the system out of equilibrium, which could also result in a neutrino isocurvature IC (see~\cite{Non_Equilibrium} and references therein for examples on non-equilibrium dynamics in the early universe). Moreover, Given the numerous possibilities that require details beyond our scope, the exact origin of these ICs is treated agnostically in this work. To make the notation brief, we will drop the superscripts $m$ from all relevant quantities, and show results for the case $m=1$. The opposite case ($m=-1$) behaves similarly up to a sign flip for the MF due to its odd parity (see eq.~\eqref{Eq:B_final}).

\subsection{Neutrino Isocurvature}
\label{Sec:ISO}
As previously mentioned, $\mathcal{V}$-modes decay with expansion of the Universe as $a^{-2}$, unless they are sourced by an initial anisotropic stress. One possibility for this to happen is if neutrinos and photons had opposite but equal vorticities initially, i.e. an Isocurvature (ISO) IC, as was considered for scalar perturbations~\cite{Ref1_AntonyPaper} in the so-called neutrino isocurvature velocity mode. 
From here, the neutrino anisotropic stress can grow and sources $\mathcal{V}$-modes through eq.~\eqref{Eq:Evol_Phi_ModeFunction}. This is why generating such ICs is quite difficult.  For more details, see Refs.~\cite{Lewis:2004kg,Itchiki_V-modes,Paoletti_theory,Rebhan:1991sr,Rebhan:1994zw}.

In order to solve the differential equations listed above, we need to specify their ICs, which correspond to the time when modes were outside the horizon. Seeking first regular solutions, then all quantities can be expanded in powers of $\eta$, since $k \eta \ll 1$ outside the horizon.  For example, the V-mode can be written as
\be
\Phi = \Phi_0+\Phi_1 \eta + \Phi_2\eta^2+ {\cal O}(\eta^3).
\label{Eq:TCA_Phi}
\ee  
where $\Phi_i$ is the value of $\Phi$ for a given mode $k$ at the $i^{\rm th}$  order in $\eta$. Moreover, tight coupling allows to consider only the common velocity $v_f$ of baryons and photons~\eqref{Eq:Def_vf}, and higher order multipoles ($\Theta_2,E_2,\Theta_3,E_3,\dots$) are suppressed. Therefore, we look for a solution for the desired quantities ($\Phi,\ v_f,\ v_\nu,\ {\cal N}_{\ell}...$) order-by-order in $\eta$. 

Before finding these ICs, we need to define a few useful relations.\footnote{Note that eqs.~\eqref{Eq:Scale_Factor} and~\eqref{Eq:1+R_TCA} apply to all ICs considered here.} First, the solution to the Friedmann equation when dark energy is negligible is (setting the scale factor today $a_0=1$)
\be
a = \HH_0^2 \Omega_m \left(\frac{\eta}{\omega}+\frac{\eta^2}{4} \right); \quad \omega = \frac{\HH_0 \Omega_m}{\sqrt{\Omega_r}},
\label{Eq:Scale_Factor}
\ee
where ${\cal H}_0$ is the current Hubble constant in conformal time and $\Omega_i$ is the current energy density of species $i$ relative to the critical density $\rho_{\rm cri}=3{\cal H}_0^2/8\pi G$ (subscripts $m$ and $r$ stand for matter and radiation, respectively). Second, starting from eq.~\eqref{Eq:GenEuler}, in the strong tight coupling regime, where there are no anisotropic stresses, $\tilde{v}_f$ satisfies
\be
\tilde v_f = \frac{\tilde{v}_0}{1+R}
\label{Eq:vf_TCA}
\ee
where $\tilde{v}_0$ is a constant to be determined. Third, using eq.~\eqref{Eq:Scale_Factor}, one can show that
\be
\frac{1}{1+R} = 1- \frac{3{\cal H}_0\Omega_b\sqrt{\Omega_r}}{4\Omega_{\gamma}}\eta+{\cal O}(\eta^2).
\label{Eq:1+R_TCA}
\ee
With these relations, applying expansion eq.~\eqref{Eq:TCA_Phi} to all required quantities, we find
\be
\VM = \Phi_0 \left(1 - c_\HH\eta\right) + {\cal O}(\eta^2),
\label{Eq:Phi_TCA}
\ee
\be
{\tilde v}_{\nu}= -\Phi_0 \left(\frac{5\Omega_r}{4\Omega_\nu} + 1\right)+ {\cal O}(\eta^2),
\label{Eq:N1_TCA}
\ee
\be
{\cal N}_2 =-\frac{k}{2\sqrt{3}} \Phi_0 \left(\frac{5}{2} 
\frac{\Omega_r}{\Omega_\nu}\eta+ c_\HH \eta^2\right)+ {\cal O}(\eta^3),
\label{Eq:N2_TCA}
\ee
and
\be
\tilde v_f = \frac{\Phi_0}{1+R}\left(\frac{5 \Omega_r +4	\Omega_\nu}{4 \Omega_\gamma}\right),
\label{Eq:vf_TCA_final}
\ee
where
\be
c_\HH = \frac{15 \Omega_m \sqrt{\Omega_r} \HH_0 }{30
	\Omega_r + 8 \Omega_\nu}\,.
\ee

We can now derive an IC for the magnetic field. In the strongly tight coupled regime, one can show that eq.~\eqref{Eq:TCV} reduces to
\be
\frac{-\tau' \delta v}{R} = \frac{\HH \tilde v_f}{1+R}\,.
\label{Eq:deltav_vf_ISO}
\ee
Inserting this into eq.~\eqref{Eq:E_useful}, using that at early times $x_e=1$, and then integrating eq.~\eqref{Eq:B_final}, we find
\be\label{Eq:B_TCA_ISO}
a^2 {\cal B} = -k \Phi_0\frac{m_p\mathcal{H}_0\sqrt{\Omega_r}}{e}\left(\frac{5\Omega_r+4\Omega_{\nu}}{4\Omega_{\gamma}}\right)\eta + {\cal O}(\eta^2)
\ee
where the integration constant was set by assuming that there is no magnetic field when $\eta\rightarrow0$.

\subsubsection{Evolution of Perturbations and Best-Fit Spectra}

In order to compare this analysis with data, we need to compute angular power spectra for the relevant quantities. We define this for the $\mathcal{V}$-modes as 
\be
\sum_{m=\pm1}\langle \VM^{(m)}_0(\gr{k}) \VM_0^{(m)\star}(\gr{k}') \rangle =
(2\pi)^3\delta(\gr{k} - \gr{k}') P_\VM(k)
\label{Eq:V_PS}
\ee
where ${}^\star$ denotes complex conjugation, $\delta(\gr{k}-\gr{k'})$ is the Dirac delta and $P_{\Phi}$ is the primordial power spectrum of $\Phi$. Using eq.~\eqref{Eq:V_PS}, we can compute CMB spectra of the form (see e.g. eq.~(56) of~\cite{HuAndWhite} for the definition)
\be
C_{\ell}^{XY} = \sigma_{\ell}\int \dd k k^2{\cal T}_X(k,\eta){\cal T}{}^\star_Y(k,\eta) P_\VM(k)
\label{Eq:CMB_Spect}
\ee
where $X,Y=\{\Theta,E,B\}$,
\be
\sigma_{\ell} = \frac{2}{\pi(2\ell+1)^2},
\label{Eq:Sigma_ell}
\ee
and ${\cal T}_{X,Y}(k,\eta)$ is the corresponding transfer function. The latter takes into account the evolution of the perturbations through eqs.~\eqref{Eq:Evol_Phi_ModeFunction},~\eqref{Eq:Hierarchy} and~\eqref{Eq:GenEuler}. The primordial $\mathcal{V}$-modes power spectrum can be defined in analogy to the scalar or tensor ones, and takes the form 
\be
P_{\Phi}=r_v A_s \bigg(\frac{k}{k_*}\bigg)^{n_v}
\label{Eq:Primodial_V_PS}
\ee
where $r_v=A_v/A_s$ is the amplitude of the $\mathcal{V}$-modes' power spectrum, $A_v$, relative to that of scalars, $A_s$ (same as $r$ for tensor modes), $n_v$ is the spectral index and $k_*$ is a pivot scale (taken to be 0.05 Mpc$^{-1}$). Therefore, in addition to the classical $\Lambda$CDM parameters\footnote{These are the baryon and CDM densities, $\omega_b$ and $\omega_c$ respectively, the angular scale of the sound horizon at recombination $\theta_s$, the amplitude and spectral index of scalar perturbations, $A_s$ and $n_s$ respectively, and the optical depth to reionization $\tau$.}, $r_v$ and $n_v$ will be the input parameters needed to study the evolution of the whole system of $\mathcal{V}$-modes, MF and the primordial plasma.

In order to solve for the spectra starting from the ICs eqs.~\eqref{Eq:Phi_TCA}-\eqref{Eq:vf_TCA_final}, we run \texttt{CLASS} providing it with values of the main cosmological parameters, i.e. the six $\Lambda$CDM ones, $r_v$ and $n_v$. We report the best-fit values of $r_v$, $n_v$ and $r$ (when applicable) in Table~\ref{Table: Best-Fits}.

\begin{figure}[!htb]
	\centering
	\includegraphics[width = 0.7\textwidth]{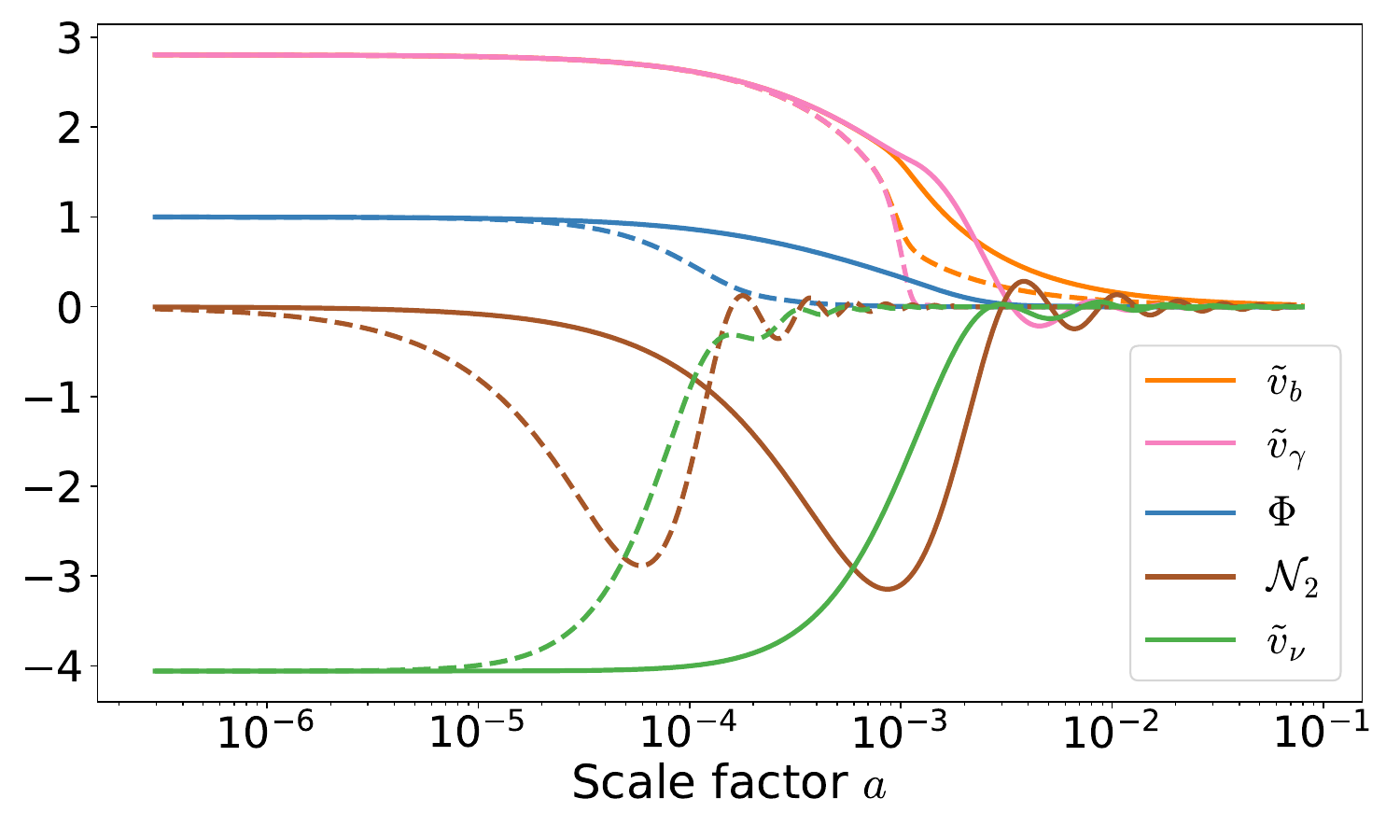}
	\caption{\justifying Evolution of $\tilde{v}_b,\ \tilde{v}_{\gamma},\ \Phi,\  \mathcal{N}_2,$ and $\tilde{v}_{\nu}$ (relative to $\Phi_0$) with the scale factor $a$. Solid lines correspond to scale $k=0.01$ Mpc$^{-1}$ while dashed ones correspond to $k=0.1$ Mpc$^{-1}$. Cosmological parameters used are from the best-fit to Planck~\cite{Planck_Primary,Planck_Lensing} SPT-3G 2018~\cite{Dutcher_etal,Balkenhol_etal} and BAO data~\cite{6dFGS,BOSS_DR12,BOSS_DR7,eBOSS_DR16}.}
	\label{Fig:ISO_Perts_vs_eta}
\end{figure}

First, let us examine the evolution of $\Phi,\ \tilde{v}_b,\ \tilde{v}_{\nu}$, $ \tilde{v}_{\gamma}$ and $\mathcal{N}_2$ shown in Figure~\ref{Fig:ISO_Perts_vs_eta}\footnote{One can compare this figure to Fig.~1 in~\cite{Lewis:2004kg}.} (relative to $\Phi_0$) as a function of $a$. At very early times, when both modes $k=0.01,0.1$ Mpc$^{-1}$ were outside the horizon, baryons and photons had the same velocities, and the balance between $\Phi,\ \tilde{v}_{f}$ and $\tilde{v}_{\nu}$ ensured vanishing neutrino quadrupole, as expected. However, since $\Phi$ deviates from a constant value as $\propto c_\HH \eta$ at early times, and since $c_\HH$ is much smaller than the mode $k$ considered, it is expected that ${\cal N}_2$, whose leading behaviour is $\propto k \eta$, varies before $\Phi$ does. This is particularly evident for the smaller scale ($k=0.1$ Mpc$^{-1}$) mode.  

Second, once a mode enters the horizon, the balance between the different quantities starts to fade as the V-mode starts decaying during radiation domination. At the same time, the neutrino quadrupole stops growing and starts oscillating. However, in contrast to the scalar perturbations case where baryons velocity increases, it decays (see eq.~\eqref{Eq:Evol_baryon}) for $\mathcal{V}$-modes. This means that after reaching equality, when baryons dominate the energy budget, they will drag the photons with them. This explains the decrease in both photon and baryon velocities seen in Figure~\ref{Fig:ISO_Perts_vs_eta}, which continues until they start separating as they approach recombination, where tight coupling is no longer applicable, and photon dipole starts oscillating.
 
\begin{figure}[!htb]
 	\centering
 	\includegraphics[width = 0.7\textwidth]{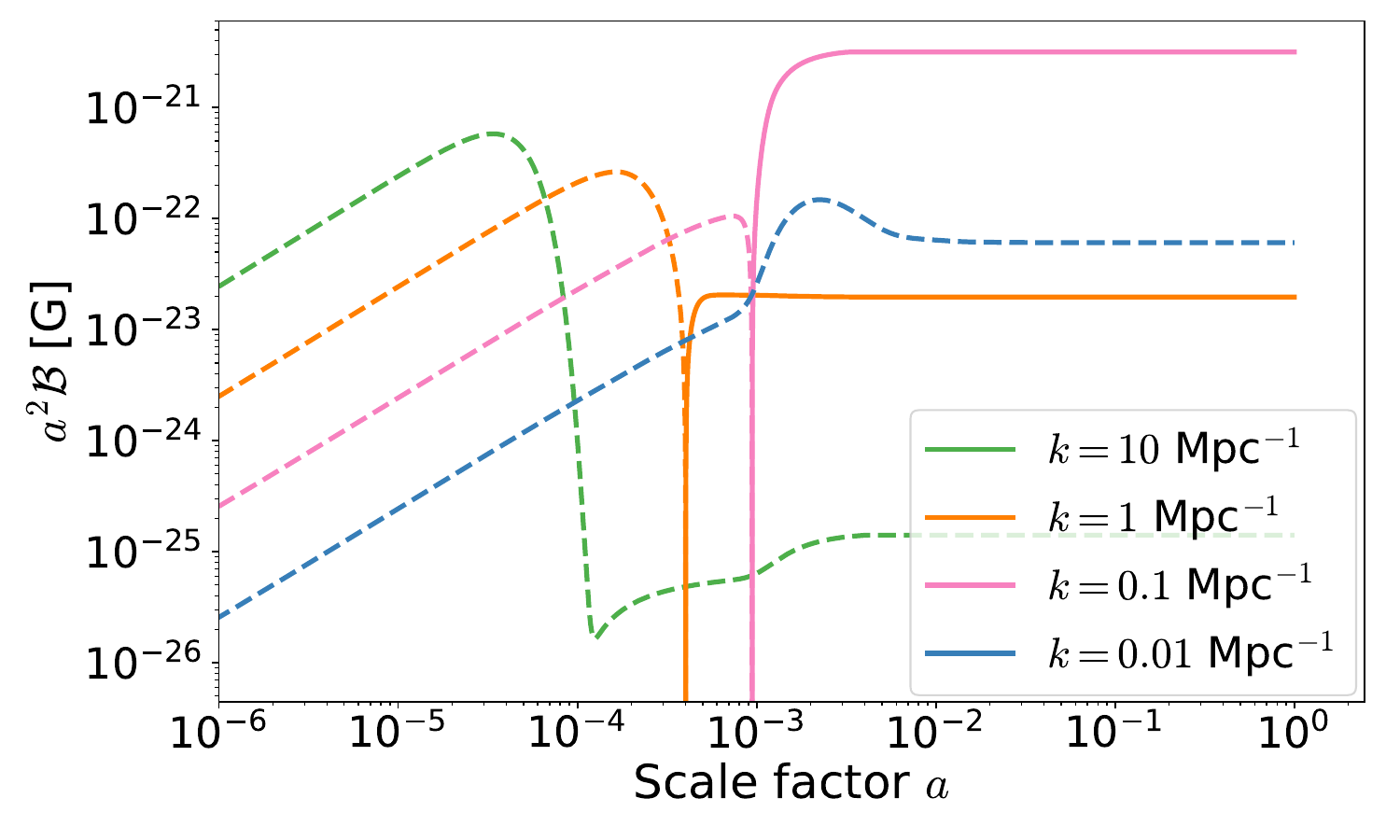}
 	\caption{\justifying Evolution of the comoving MF $a^2\mathcal{B}$ (relative to $\Phi_0$) with the scale factor $a$ for the four modes listed in the legend. Dashed lines correspond to negative values. Cosmological parameters used are from the best-fit to Planck~\cite{Planck_Primary,Planck_Lensing} SPT-3G 2018~\cite{Dutcher_etal,Balkenhol_etal} and BAO data~\cite{6dFGS,BOSS_DR12,BOSS_DR7,eBOSS_DR16}.}
 	\label{Fig:ISO_B_vs_a}
 \end{figure}

 We show the evolution of the MF produced by $\delta v$ in Figure~\ref{Fig:ISO_B_vs_a}. First, the vertical shifts between the different curves are proportional to $k$ as anticipated in the approximation \eqref{Eq:B_TCA_ISO}. Second, for small scales ($k\gtrsim0.1$ Mpc$^{-1}$), $|a^2\mathcal{B}|$ increases linearly from early times and reaches a maximum value slightly before the onset of Silk damping~\cite{Silk_Damping}\footnote{Recall that, for a given mode $k$, the scale factor at which Silk damping onsets is $a_{\text{damp}}\sim 10^{-4} k^{-2/3}$(see eq. (B-16) of~\cite{Hu_Sugiyama}).}. This is expected since this is when $|\delta v|$ reaches its maximum value and then starts decreasing (see eqs.~\eqref{Eq:deltav_vf_ISO} and~\eqref{Eq:B_TCA_ISO}). During radiation domination, there's a slight positive difference between $v_{\gamma}$ and $v_b$, as photons drag the baryons along with them. However, once diffusion starts, photon perturbations start diminishing, and baryons then start having a larger velocity, i.e. $\delta v>0$. This can be corroborated with the $\tilde{v}_{\gamma}$ (pink) and $\tilde{v}_b$ (orange) dashed curves of Figure~\ref{Fig:ISO_Perts_vs_eta} for the $k=0.1$ Mpc$^{-1}$ case in Figure~\ref{Fig:ISO_B_vs_a}. This explains why the magnetic field changes sign (negative values are represented by dashed line in Figure~\ref{Fig:ISO_B_vs_a}) for those modes, after which its amplitudes simply asymptotes at the last value it reached before diffusion washes out all perturbations at the corresponding scale. 
 
 On the other hand, large scale modes ($k=0.01$ Mpc$^{-1}$) remain intact, as they spent most of the time outside the horizon. Moreover, around the time of photon decoupling, there's a small boost in $\tilde{v}_{\gamma}$ compared to $\tilde{v}_b$ (Figure~\ref{Fig:ISO_Perts_vs_eta}), which explains the observed bump in the $k=0.01$ Mpc $^{-1}$ curve of Figure~\ref{Fig:ISO_B_vs_a}. This protects $a^2\mathcal{B}$ from changing sign when diffusion kicks-in, after which it asymptotes. 
 
 Lastly, and most importantly, the amplitude of the MF on all relevant scales is too small to be the $B_{\mathrm{seed}}$ needed to explain all astrophysical MFs observed. As mentioned in the Introduction~\ref{Sec:Intro}, $B_{\text{seed}}\sim\mathcal{O}(10^{-14})$ G at redshifts $z\sim 0-2$ is needed to explain galactic, inter-galactic and inter-cluster MFs, which is not the case of PMFs produced from $\mathcal{V}$-modes with ISO ICs.  One can quantify the MF's amplitude, as done in previous works~\cite{Planck:2015,Kosowsky_etal:2002}, as the average of its power spectrum today smoothed with a Gaussian on a scale $\lambda_1=1$ Mpc: 
 
 \be
 \mathcal{B}_{1}^2 (\eta_0)= \frac{1}{2\pi^2}\int \dd k k^2P_{\mathcal{B}}(\eta_0,k)e^{-\lambda_1^2 k^2} ,
 \label{Eq:Smoothed_B}
 \ee 
 where $\eta_0$ is conformal time today and 
 \be
 P_{\mathcal{B}}(\eta,k)= |\mathcal{T}_{\mathcal{B}}(\eta,k)|^2P_{\Phi}(k)
 \label{Eq:B_PS}
 \ee
 is the PMF's power spectrum (similarly to eq.~\eqref{Eq:CMB_Spect}), given in terms of its transfer function, $\mathcal{T}_{\mathcal{B}}$, and the primordial V-mode spectrum, eq.~\eqref{Eq:Primodial_V_PS}. With this definition, we find ${\cal B}_1 = 5.07\times10^{-27}$ G, still too small to act as $B_{\text{seed}}$. This means that the associated energy density is subdominant compared to the one of the photon background since 
 \be\label{Eq:ratiorho}
\frac{\rho_B}{\bar{\rho}_\gamma} \simeq 10^{-41} \times \left(\frac{\mathcal{B}_{1}}{10^{-26}{\rm G}}\right)^2\,.
 \ee
 where $\rho_B = \mathcal{B}_{1}^2/2$ is the energy density of the magnetic field. Since the anisotropic stress of the PMF is of the order of $\rho_B$, the PMF will not back-react on the CMB and modify its power spectra. This is true for both the helical and non-helical parts of the PMF (see eqs.(3.20) and (3.21) of~\cite{Helicity_CMB4}). We will see that the two other ICs  produce a similar amplitude of PMFs, and will not back-react on the CMB either.
 
\begin{figure}[!htb]
	\centering
	\includegraphics[width = 0.7\textwidth]{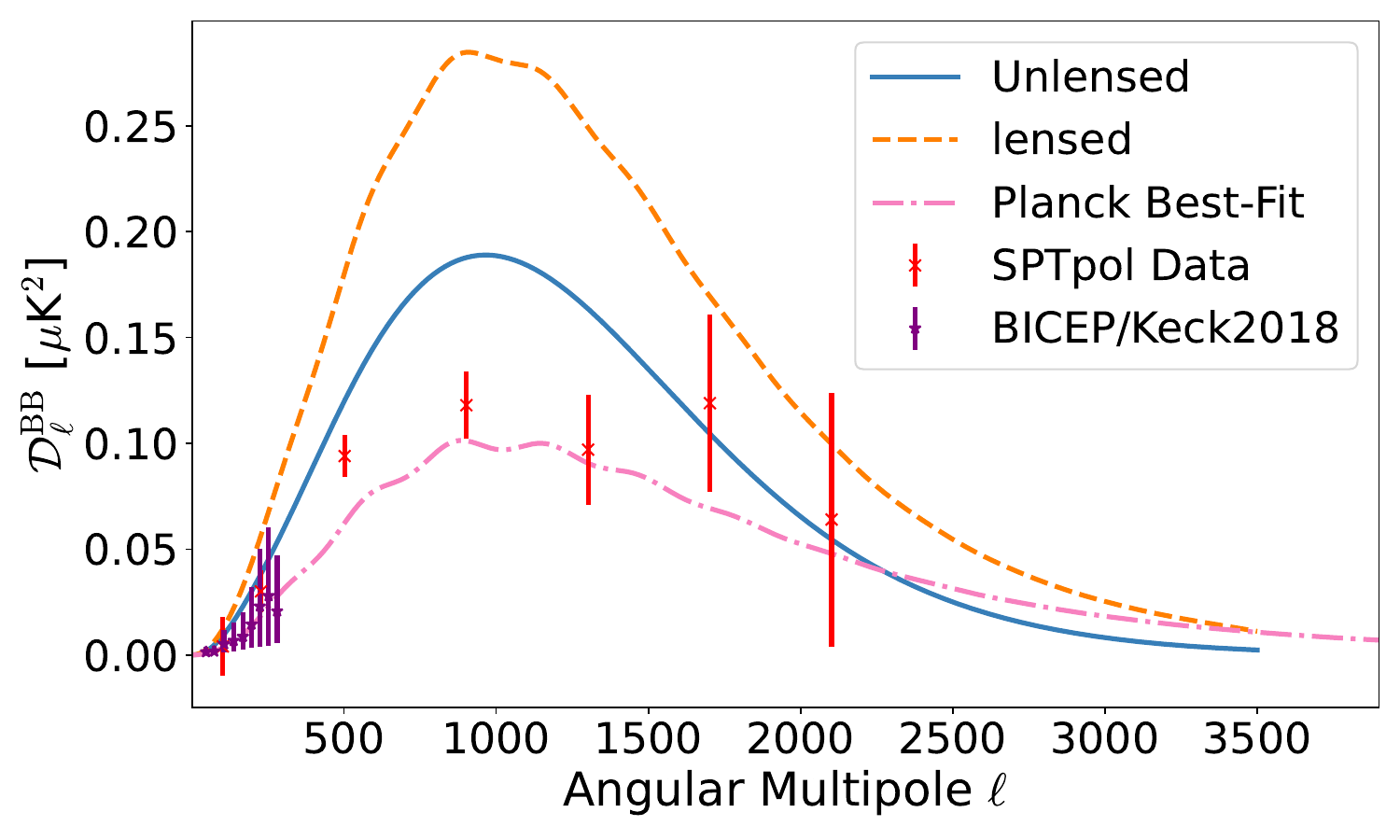}
	\caption{\justifying Best-fit power spectrum of CMB polarization $B$-modes for the case of ISO ICs compared to data. The latter includes data points from BICEP/Keck~\cite{BICEP_Keck} (purple star-shaped) and SPTpol~\cite{SPTpol} (red x-shaped). The solid blue curve is the unlensed $BB$ spectrum coming purely from $\mathcal{V}$-modes, while the dashed orange curve is for the total lensed one. The pink dot-dash curve is the Planck $\Lambda$CDM best-fit lensed spectrum.}
	\label{Fig:ISO_BB_lin}
\end{figure}

Moreover, in Figure~\ref{Fig:ISO_BB_lin}, we plot the best-fit power spectrum of the CMB's polarization $B$-modes, and contrast it with data points from BICEP/Keck~\cite{BICEP_Keck} and SPTpol~\cite{SPTpol}. These two data sets are complementary, as the former covers small $\ell$ ranges, while the latter intermediate ones, providing valuable constraints when combined. It is evident that the $B$-modes spectrum resulting purely from $\mathcal{V}$-modes differs from that of tensor modes as the former dominates at intermediate scales, while tensor modes do not (see Figure~\ref{Fig:CMB_Spectra} for a comparison). It is also apparent from Figure~\ref{Fig:ISO_BB_lin} that the $\Lambda$CDM Planck best-fit spectrum (pink dot-dash curve), which is equivalent to lensed $EE$ spectrum, is a better match to the data compared to the corresponding one from $\mathcal{V}$-modes with ISO ICs (orange dashed curve). 

The above outcomes for $\mathcal{B}$ and the $B$-modes spectra are for a best-fit value $r_v=1.2\times10^{-3}$, as shown in Table~\ref{Table: Best-Fits}. In an attempt to increase the amplitude $\mathcal{B}$, we searched for the best-fit of the other cosmological parameters after setting $r_v=0.01$. Not only the resultant $\mathcal{B}$ is still too small to explain every observable MF (${\cal B}_1 = 4.12\times10^{-26}$ G), but also the $B$-modes spectrum is even a worse fit to BICEP/Keck and SPTpol data. Moreover, we have also checked that including tensor modes in addition to $\mathcal{V}$-modes worsens the fit even more than the previous case, with the resulting best-fit amplitudes given in the 4$^{\text{th}}$ row of Table~\ref{Table: Best-Fits}. Both tensor and vector modes produce not only $B$-modes, but also polarization $E$-modes. This means that the final lensed $BB$ spectrum, which is what should be compared with data, will have almost double the power compared to just $\mathcal{V}$-modes. Since including tensor modes with $\mathcal{V}$-modes is more realistic, this result disfavours the ISO IC even more.

\textbf{Summary:} Assuming an Isocurvature IC between neutrino and photon velocities for $\mathcal{V}$-modes in the primordial Universe cannot produce large enough seed MFs that could explain observations. However, even if a theoretical motivation is still needed, these $\mathcal{V}$-modes might still exist and an upper bound on their amplitude should be set. Specifically, including data from BICEP/Keck~\cite{BICEP_Keck} and SPTpol~\cite{SPTpol} is crucial to that, which will be the topic of our future work. 

\subsection{Neutrino Octupole}

The analytical description of the initial behaviour provided in the previous section relies on truncating the photon and neutrino hierarchies at the quadrupole level ($\ell=2$). Although it is a good approximation, it would be a natural generalization to relax this assumption and allow the next multipole, specifically the neutrino octupole ($\mathcal{N}_3$), to have a non-vanishing IC, which we abbreviate as the OCT IC. This is highly unconventional, and an origin for this IC is still to be determined.

To understand the details of the dynamics, let us look at the evolution equations at play. On the neutrino side, from eq.~\eqref{Eq:Nu_Hierarchy}, we get

\be
\mathcal{N}_1 {}' = -\frac{\sqrt{3}}{5}k\mathcal{N}_2 + \Phi{}'
\label{Eq:Nu_dipole}
\ee
for the neutrino dipole ($\ell=1$),
\be
\mathcal{N}_2 {}' = k\bigg[\frac{1}{\sqrt{3}}\mathcal{N}_1 - \frac{\sqrt{8}}{7}\mathcal{N}_3\bigg]
\label{Eq:Nu_quad}
\ee
for its quadrupole ($\ell=2$), and

\be
\mathcal{N}_3 {}' = k\frac{\sqrt{8}}{5}\mathcal{N}_2
\label{Eq:Nu_oct}
\ee
for the octupole ($\ell=3$) when ${\cal N}_4$ is negligible. Substituting eq.~\eqref{Eq:Nu_oct} in eq.~\eqref{Eq:Nu_dipole} for $\mathcal{N}_2$ and integrating, we find
\be
\mathcal{N}_1 = -\sqrt{\frac{3}{8}}\mathcal{N}_3 +\Phi +C,
\label{Eq:Sol_N1}
\ee
where $C$ is an integration constant. The latter has to be set such that there's no neutrino dipole in the limit $\eta\rightarrow0$. Note also that the absence of $k$ means that this relationship will be maintained through all scales in the very early Universe, unlike the ISO case. 

On the other hand, inserting eq.~\eqref{Eq:Sol_N1} in eq.~\eqref{Eq:Nu_quad}, the neutrino quadrupole's evolution takes the form
\be
\mathcal{N}_2{}' = -k \bigg[\frac{15}{7\sqrt{8}}\mathcal{N}_3 -\frac{1}{\sqrt{3}}\big(\Phi+C\big)\bigg],
\label{Eq:First_deriv_N2}
\ee
where the influence of the octupole is now evident. One can actually derive another evolution equation, by
taking the time derivative of eq.~\eqref{Eq:Nu_quad} and inserting eqs.~\eqref{Eq:Nu_dipole} and~\eqref{Eq:Nu_oct}, to get

\be
\mathcal{N}_2 {}'' +  \frac{3k^2}{7}\mathcal{N}_2 = \frac{k}{\sqrt{3}}\Phi{}',
\label{Eq:Evol_Eq_N2}
\ee
which highlights the oscillatory nature of the quadrupole, with a force term given by $\Phi{}'$. The latter is given by eq~\eqref{Eq:Evol_Phi_ModeFunction}, which we write explicitly as

\be
\Phi{}' + 2\mathcal{H}\Phi = -\frac{8\sqrt{3}\mathcal{H}_0^2}{5a^2k}\big(\Omega_{\gamma}\Theta_2+\Omega_{\nu}\mathcal{N}_2\big),
\label{Eq:Evol_Eq_Phi}
\ee
where we used the definition of the anisotropic stress eq.~\eqref{Eq:Pim_def} and $\Omega_i(a)=\Omega_ia^{-4}$ for $i=\gamma,\nu$.

The last ingredients we need to complete the quantitative description is for the photo-baryonic fluid. We use eqs.~\eqref{Eq:Evol_Photon_Dipole} and~\eqref{Eq:Evol_baryon} for the photon and baryon velocities' evolution, respectively, and 

\be
\Theta_2{}' = \frac{1}{\sqrt{3}}k\Theta_1 + \frac{\tau{}'}{10}\big(\Theta_2-\sqrt{6}E_2\big) - \tau{}'\Theta_2
\label{Eq:Evol_G_Quad}
\ee
for the photon quadrupole's evolution ($\ell=2$ of eq.~\eqref{Eq:Hierarchy}), neglecting the octupole contribution $\Theta_3$. 

Qualitatively, these equations tell us that having a non-trivial neutrino octupole initially sources the evolution of its quadrupole (eq.~\eqref{Eq:Nu_quad}) and then its dipole (eq.~\eqref{Eq:Nu_dipole}). The former then sources the $\mathcal{V}$-modes' evolution (eq.~\eqref{Eq:Evol_Eq_Phi}), hence sustaining its constancy, but the gradual deviation from constancy is such that eq.~\eqref{Eq:Sol_N1} must be satisfied on large scales. On the photon side, when using the TCA, this results in $\Theta_1\approx\Phi$ if we set $\tilde{v}_f=0$ initially since there are no isocurvature initial conditions in that case (eq.~\eqref{Eq:Euler_Photon}). Then, this affects photon diffusion, i.e. anisotropic stress (eq.~\eqref{Eq:Evol_G_Quad}), resulting in an increase in the velocity difference between photons and baryons (eqs.~\eqref{Eq:Evol_Photon_Dipole} and~\eqref{Eq:Evol_baryon}). This difference is what creates the electric field (eq.~\eqref{Eq:E_useful}) needed to generate the PMF (eq.~\eqref{Eq:B_final}). Given how indirect the production of PMFs is in this IC, we expect a smaller photon-baryon velocity difference, and hence a smaller amplitude of PMFs, compared to the ISO case, for the same primordial $\mathcal{V}$-modes power spectrum. 

After applying the approximations eq.~\eqref{Eq:TCA_Phi}-\eqref{Eq:1+R_TCA} to all the relevant elements, we can solve the system of coupled differential equations, eqs.~\eqref{Eq:Nu_dipole}-\eqref{Eq:Evol_G_Quad}, \eqref{Eq:Evol_Photon_Dipole} and~\eqref{Eq:Evol_baryon}. Up to next-to-leading order (NLO) in $\eta$, we find 
\be
\Phi = \Phi_0 \left(1 -  c_\HH \eta\right),
\label{Eq:IC_Phi_OCT}
\ee

\be
\tilde{v}_{\nu} = \Phi_0 \frac{\Omega_r}{8\Omega_{\nu}}(k\eta)^2 + {\cal O}(\eta^3),
\label{Eq:IC_vnu_OCT}
\ee

\be
{\cal N}_2 =-\sqrt{3} k\Phi_0  \left(\frac{5}{12} 
\frac{\Omega_r}{\Omega_\nu}\eta+\frac{1}{6} c_\HH \eta^2\right),
\label{Eq:IC_N2_OCT}
\ee
\be
{\cal N}_3 = \Phi_0\frac{7}{2\sqrt{6}}
\left(\frac{5\Omega_r+4\Omega_\nu}{4\Omega_\nu}\right) + {\cal O}(\eta^2)\,,
\label{Eq:IC_N3_OCT}
\ee

\be
\tilde{v}_f= {\cal O}(\eta^3)\,.
\label{Eq:IC_vf_OCT}
\ee

\begin{figure}[!htb]
		\begin{subfigure}[t]{0.53\textwidth}
			\hspace{-1.8cm}
			\centering
			\includegraphics[width=\textwidth]{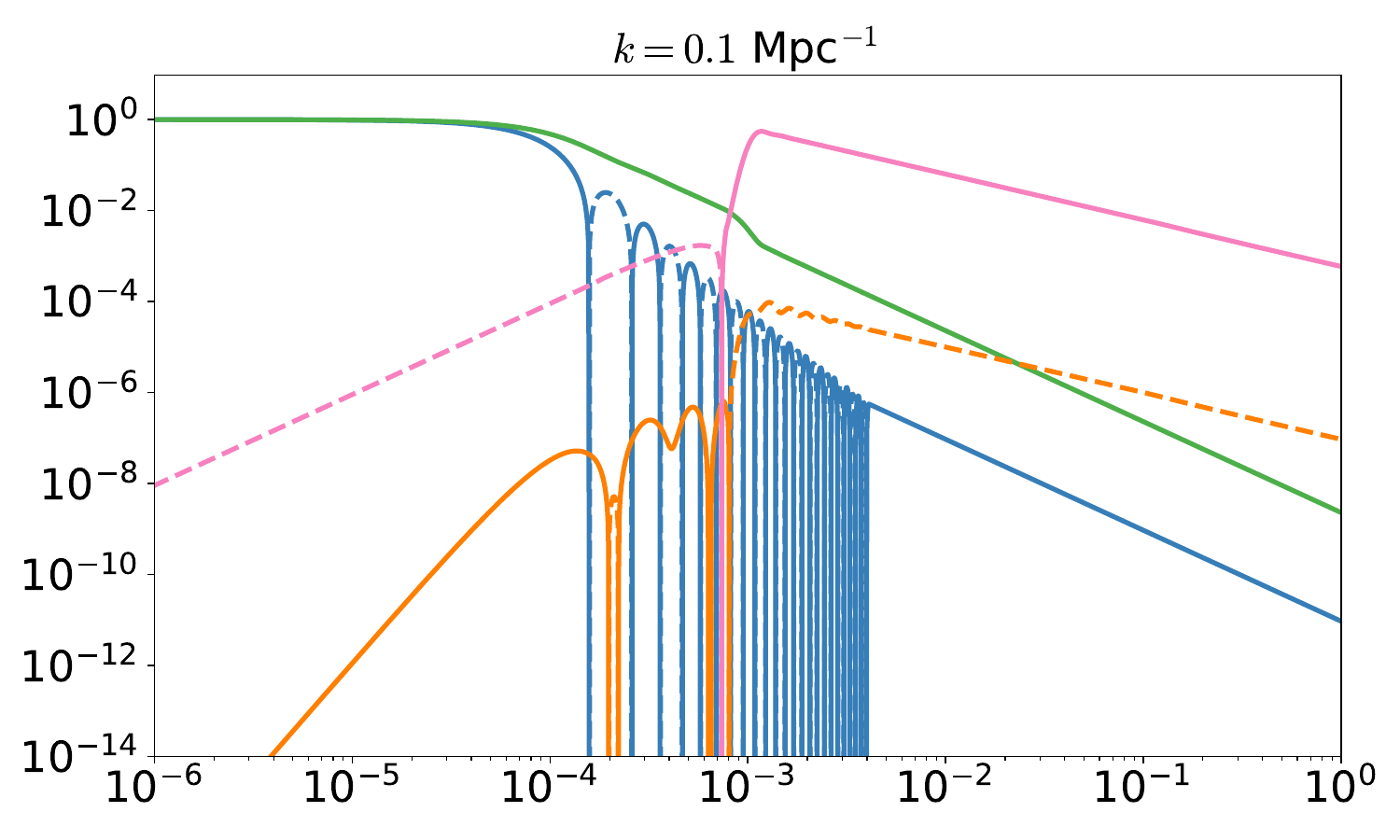}
		\end{subfigure}%
		\hfill
		\begin{subfigure}[t]{0.53\textwidth}
			\hspace{-1.8cm}
			\centering
			\includegraphics[width=\textwidth]{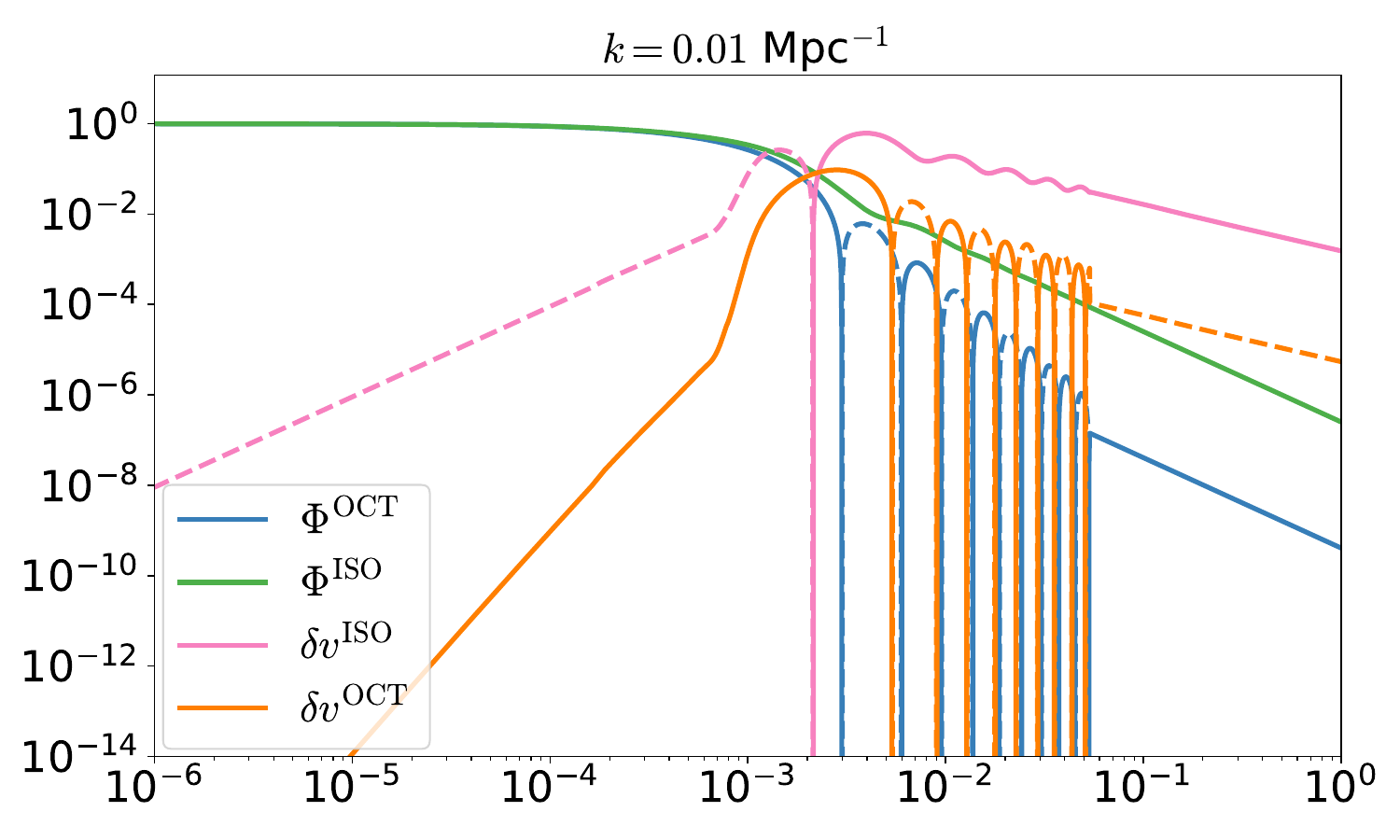}
		\end{subfigure}
		
		\begin{subfigure}[t]{0.53\textwidth}
			\hspace{-1.8cm}
			\centering
			\includegraphics[width=\textwidth]{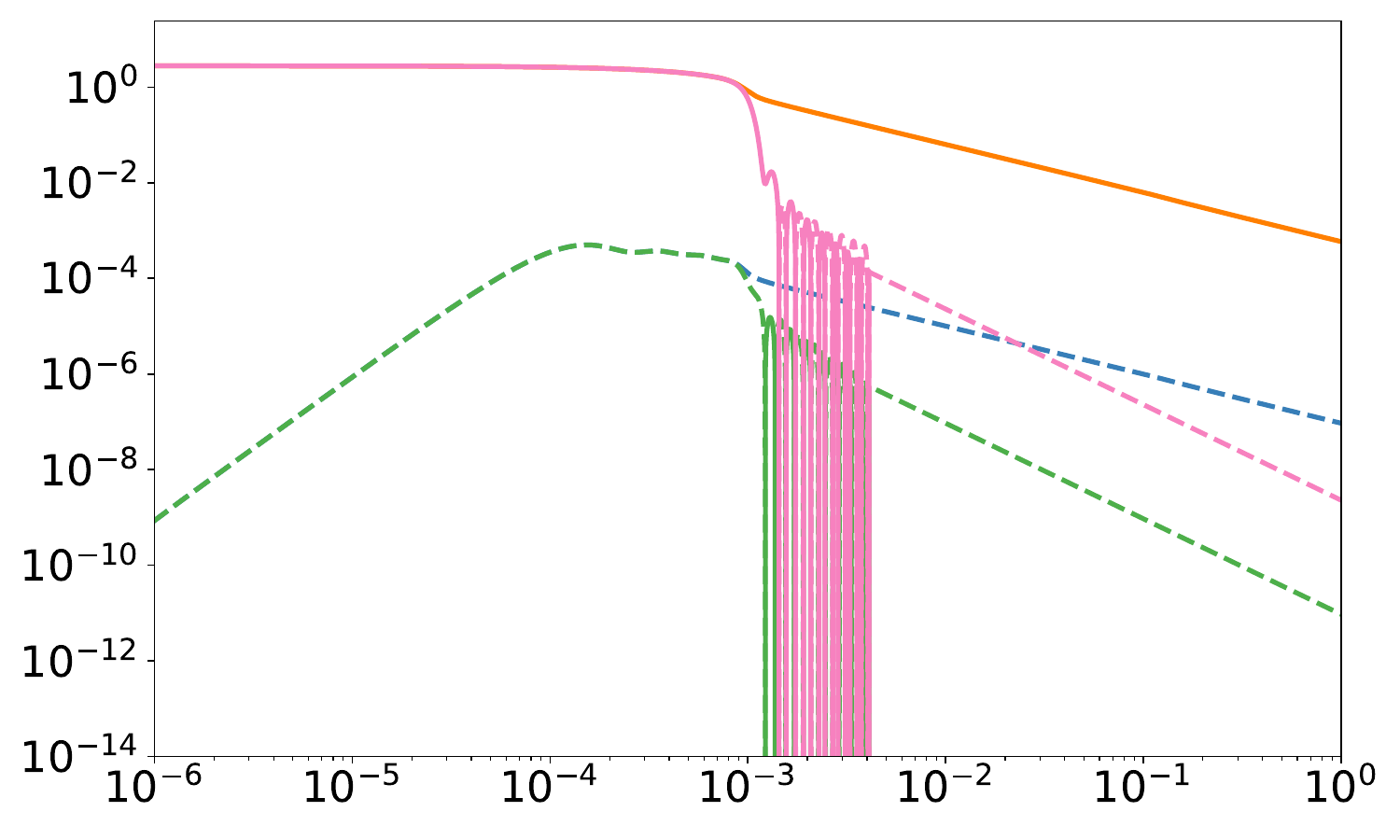}
		\end{subfigure}%
		\hfill
		\begin{subfigure}[t]{0.53\textwidth}
			\hspace{-1.8cm}
			\centering
			\includegraphics[width=\textwidth]{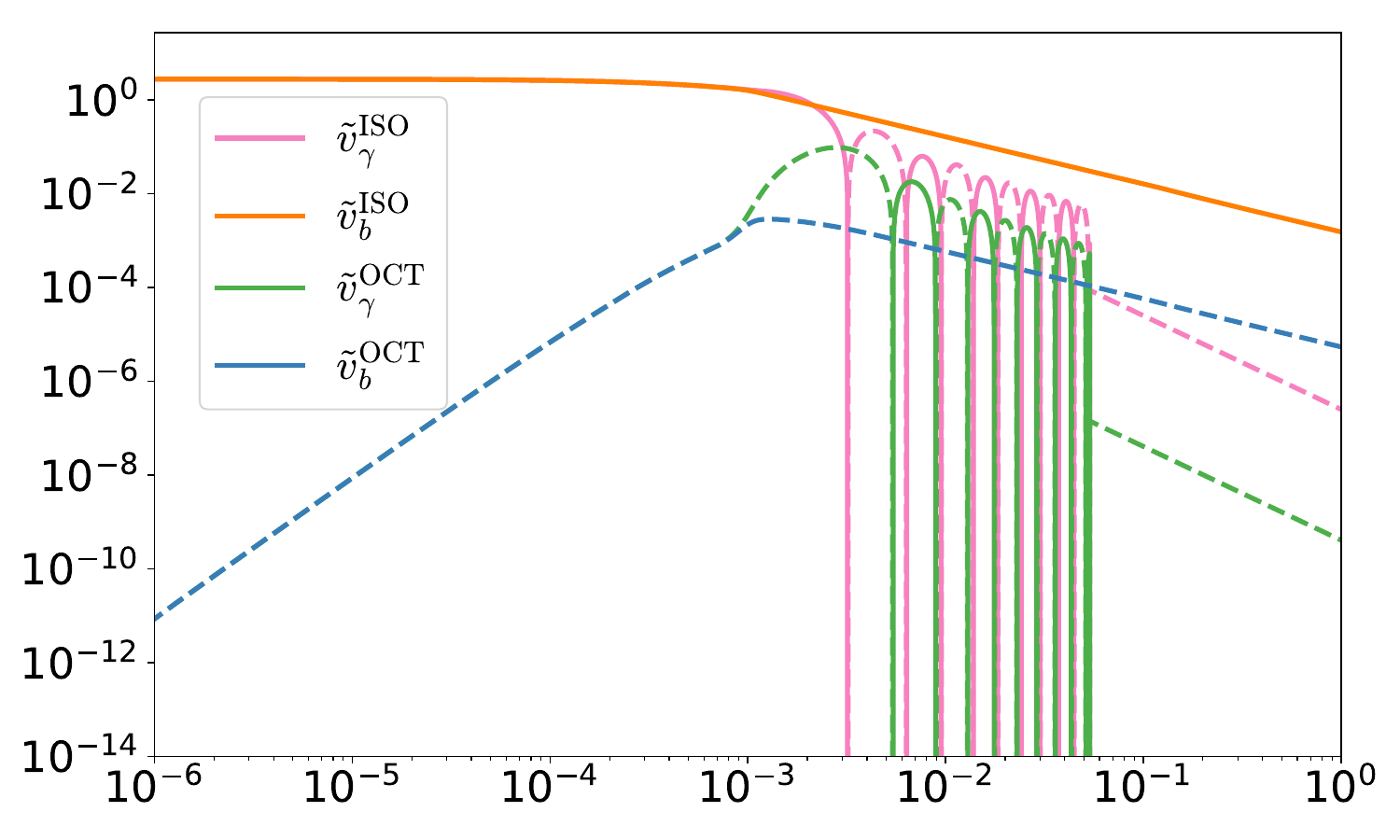}
		\end{subfigure}
		\begin{subfigure}[t]{0.53\textwidth}
			\hspace{-1.8cm}
			\centering
			\includegraphics[width=\textwidth]{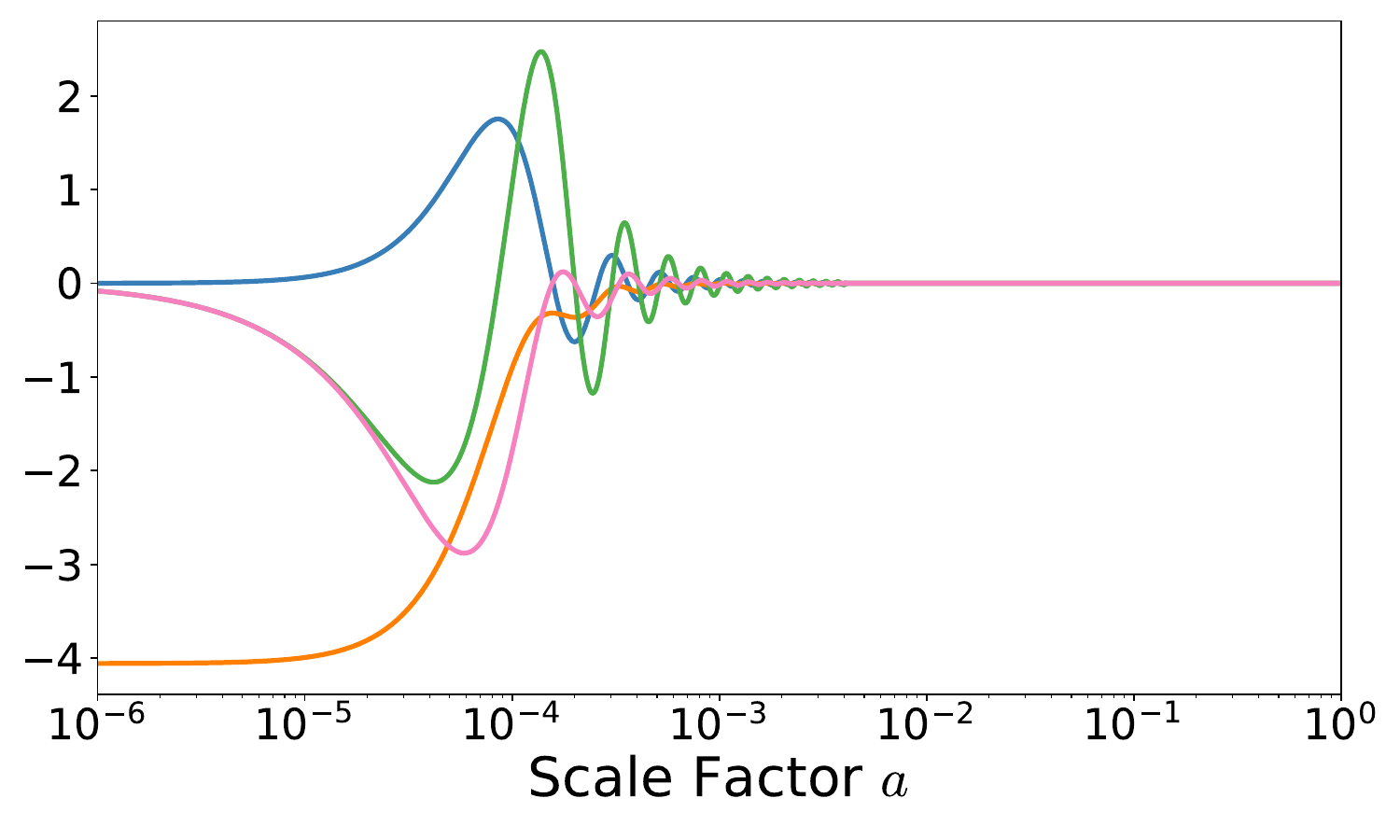}
		\end{subfigure}%
		\hfill
		\begin{subfigure}[t]{0.53\textwidth}
			\hspace{-1.8cm}
			\centering
			\includegraphics[width=\textwidth]{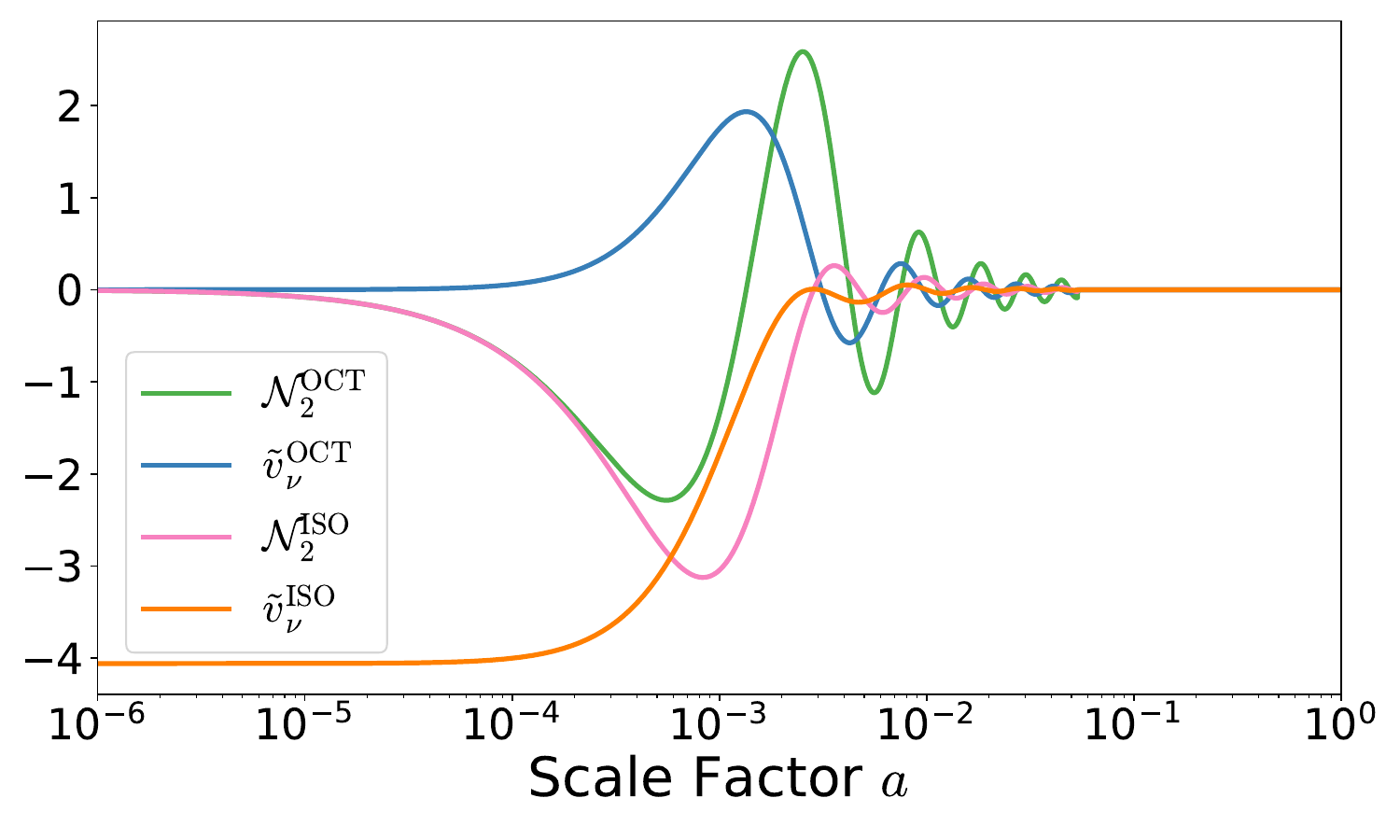}
		\end{subfigure}
		\caption{\justifying Evolution of $\Phi,\ \delta v,\ \tilde{v}_b,\ \tilde{v}_{\gamma},\ \tilde{v}_{\nu} $ and $\mathcal{N}_2$ (relative to $\Phi_0$) with the scale factor $a$ for neutrino OCT IC. Each term is plotted in comparison with the corresponding term for ISO IC. The plots on the left are for $k = 0.1$ Mpc$^{-1}$, while the ones on the right are for $k = 0.01$ Mpc$^{-1}$. The dashed lines in the top and middle plots correspond to negative values, while the solid lines to positive ones. The bottom plots are presented in semi-log scales to ease visualisation. In producing these plots, we set $r_v=10^{3}$, $n_v=0$ and the Planck best-fit for the $\Lambda$CDM cosmological parameters. In the top and middle plots, a radiation streaming approximation (RSA)~\cite{CLASSII} is applied when oscillations become large at later times. This is equivalent to setting $\Theta_\ell={\cal N}_\ell=0$ for $\ell \geq 1$, which means in particular that $\tilde{v}_\gamma= - \Phi$.}
\label{Fig:OCT_vs_ISO}
	\end{figure}

To better understand the mechanism at hand, we plot in Figure~\ref{Fig:OCT_vs_ISO} the evolution of the above quantities (relative to $\Phi_0$) as a function of the scale factor $a$. Starting with the V-mode (top plots of Figure~\ref{Fig:OCT_vs_ISO}), we can see that it displays a clear oscillatory behavior after the mode crosses the horizon, unlike the ISO case. This is expected given the strong oscillatory nature of the neutrino quadrupole due to the presence of a non-vanishing octupole.

 Second, after crossing the horizon, the $\mathcal{V}$-modes' amplitude is smaller compared to the ISO one. This is because part of the $\mathcal{V}$-modes is being shared by the neutrino octupole along with the dipole, as can be seen from eq.~\eqref{Eq:Sol_N1}. Finally, when a mode enters the horizon prior to photon decoupling, it suffers more from diffusion damping. This is why the small scale mode ($k=0.1$ Mpc$^{-1}$) features more oscillations and damping compared to the larger mode ($k=0.01$ Mpc$^{-1}$).

We now move to the photo-baryonic fluid. Notice first the difference in amplitude and slope of the velocities between the two ICs, as seen from the middle plots of Figure~\ref{Fig:OCT_vs_ISO}. The smallness of the amplitude is again a manifestation of the indirect sourcing of velocities with a neutrino octupole (eqs.~\eqref{Eq:Evol_Photon_Dipole}-\eqref{Eq:Evol_baryon}). On the other hand, after inserting eq.~\eqref{Eq:1+R_TCA} into eqs.~\eqref{Eq:vf_TCA_final} for the ISO IC, we can see that $\tilde{v}_f$ will have a constant term, whereas in the OCT case, $\tilde{v}_f=\mathcal{O}(\eta^3)$. This explains the slope difference between the two ICs. Moreover, even though photons and baryons are tightly coupled  originally, their velocity difference is not exactly zero. In fact, $\delta v$ initially has an oscillatory behavior, which is particularly apparent for the small scale case (top left plot of Figure~\ref{Fig:OCT_vs_ISO}). This behavior is due to the relation between $\delta v$ and $\Phi$, which will become clear shortly. Then, after decoupling, the evolution of $\delta v$ becomes dominated by $\tilde{v}_b$, which can be seen by comparing the top and middle plots of Figure~\ref{Fig:OCT_vs_ISO}.

After photon diffusion dominates the dynamics (when the baryon and photon velocities start separating), different behaviors appear at small and large scales. In the former case ($k=0.1$ Mpc$^{-1}$), for both ICs, photon velocities manifest almost the same rapid decaying oscillations, and baryons' velocity simply decays as $ a^{-1}$. The large scale ($k=0.01$ Mpc$^{-1}$) case suffers instead less damping since it enters the horizon later. Moreover, there's a substantial increase in photon velocities at the onset of diffusion for the case $k=0.01$ Mpc$^{-1}$, which is absent in the $k=0.1$ Mpc$^{-1}$ one. For the large scale mode, this bump is due to the increase in neutrino quadrupole amplitude that happens around the same time, which is transferred to the $\tilde{v}_\gamma$ through the $\mathcal{V}$-modes as can be seen from the bottom right plot of Figure~\ref{Fig:OCT_vs_ISO}. On the other hand, this increase happens much earlier for the small scale mode (bottom left plot of Figure~\ref{Fig:OCT_vs_ISO}), when photons and baryons are still tightly coupled, resulting in a barely noticeable bump in the photo-baryonic speed around that time (middle left plot of Figure~\ref{Fig:OCT_vs_ISO}). Note that there is a similar increase for the ISO case, as we saw in Figure~\ref{Fig:ISO_Perts_vs_eta}, but it is much smaller since the corresponding increase in $\mathcal{N}_2$ is (see pink dashed curve of the bottom right plot in Figure~\ref{Fig:OCT_vs_ISO}).

Lastly, a few words on the neutrino side of the dynamics, shown in the bottom plots of Figure~\ref{Fig:OCT_vs_ISO}. First, if we compare small to large scale evolution, we can see that they both show the same patterns for both ICs. The main difference comes from the small scale mode ($k=0.1$ Mpc$^{-1}$) entering the horizon earlier, resulting in it suffering more damping once neutrinos free-stream\footnote{Recall that, for a mode $k$, the scale factor at which neutrinos free-stream is $a_{\text{fs}}\sim 10^{-5} k^{-1}$(see eq.(8.83) of~\cite{Dodelson})}, while this happens later for the large scale mode. However, there's a clear difference in the dipole evolution between the two ICs. For the ISO case, the neutrino dipole follows the evolution of its main source, the V-mode, maintaining a balance with the photon dipole. On the other hand, for the OCT IC, the oscillatory behavior of the neutrino dipole follows that of the quadrupole, which itself is sourced by the octupole. This also explains why in the current case the neutrino quadrupole features higher oscillation amplitudes compared to the ISO IC. 

Before looking at the best-fit evolution of the MF and CMB $BB$ spectrum, we derive an expression for the MF's IC. As we saw from eqs.~\eqref{Eq:IC_Phi_OCT}-\eqref{Eq:IC_vf_OCT}, we need to go to NLO in order to get non-trivial expressions for the ICs. At that order, the following approximations hold\footnote{The first approximation follows from the fact that $\tilde{\Theta}_1\approx0$ at leading order in TCA, while the second one can be derived from combining the evolution equations of $\Theta_2$ and $E_2$ (neglecting their time derivatives w.r.t the other terms).}
\be
\Theta_1\approx\Phi; \ \Theta_2 \approx \frac{4k}{3\sqrt{3}\tau{}'}\Theta_1.
\label{Eq:NLO_approx}
\ee
In this case, eq.~\eqref{Eq:TCV} reduces to
\be
- \frac{\tau' \delta v}{R} = -\frac{\sqrt{3}}{5}k\left(\HH\int d\eta\Theta_2 +\Theta_2\right).
\label{Eq:deltav_vf_OCT}
\ee
These expressions show the relation between $\delta v$ and $\Phi$ (through $\Theta_2$), explaining the oscillatory behavior of $\delta v$ seen in the top plots of Figure~\ref{Fig:OCT_vs_ISO}.  Inserting this into eq.~\eqref{Eq:E_useful}, the latter into eq.~\eqref{Eq:B_final} and then integrating, we find

\be
a^2{\cal B} = A\Phi_0k^3\eta^4,
\label{Eq:B_TCA_OCT}
\ee
where 
\be
A = \frac{16\pi Gm_p^2\HH_0\Omega_r^{3/2}}{135\Omega_b\sigma_Te},
\ee
and the integration constant has been set such that there is no other magnetogenesis mechanism in the limit $\eta\rightarrow0$.

\subsubsection{Best-Fit MF and CMB $BB$ spectrum}

As done for the ISO IC, we find the best-fit parameters to data from Planck~\cite{Planck_Primary,Planck_Lensing} SPT-3G 2018~\cite{Dutcher_etal,Balkenhol_etal} and SDSS-BAO data~\cite{6dFGS,BOSS_DR12,BOSS_DR7,eBOSS_DR16}. Specifically for the $\mathcal{V}$-modes parameters, we find $r_v=15.3$ and $n_v=3.7$ for octupole ICs, as shown in the second row of Table~\ref{Table: Best-Fits}. This means that the data can accommodate a much larger amplitude of $\mathcal{V}$-modes and more blue-shifted spectrum compared to the ISO case. Nevertheless, the resultant amplitude using eq.~\eqref{Eq:Smoothed_B} is ${\cal B}_1 = 1.67\times10^{-27}$ G, which is still too small to explain all MFs observed astrophysically.

\begin{figure}[!htb]
	\centering
 	\includegraphics[width = 0.7\textwidth]{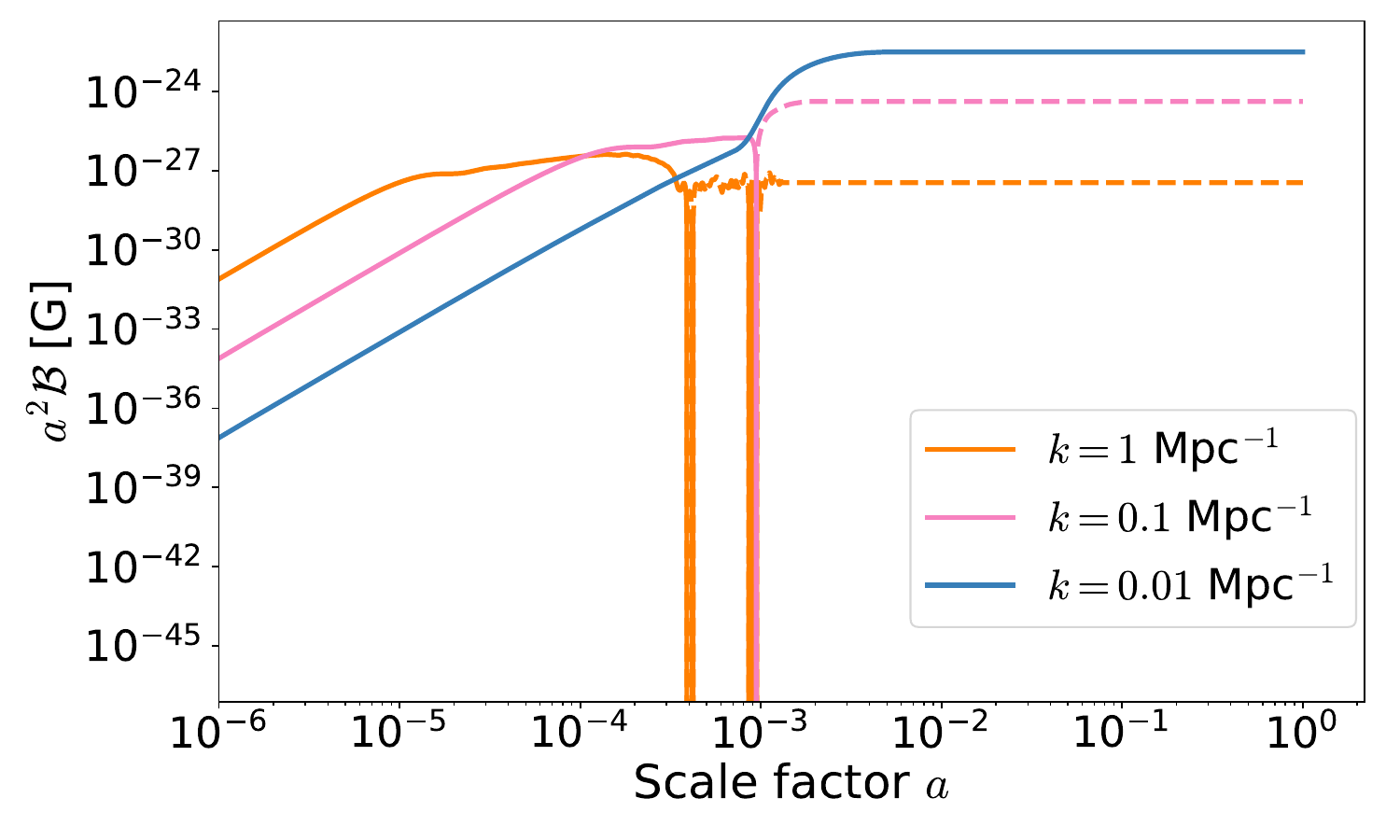}
 	\caption{\justifying Same as Figure~\ref{Fig:ISO_B_vs_a} for the OCT IC, excluding the $k=10$ Mpc$^{-1}$ mode.}
 	\label{Fig:OCT_B_vs_a}
 \end{figure}

In Figure~\ref{Fig:OCT_B_vs_a}, we show the evolution of the comoving amplitude of the MF for the three modes $k=0.01,\ 0.1$ and 1 Mpc$^{-1}.$\footnote{Note that the RSA applied in Figure~\ref{Fig:OCT_vs_ISO} does not affect this result, since the RSA is applied after the MF freezes.} First, we can see from this figure that the MF today is $\sim$ 3-7 orders of magnitude smaller compared to the ISO case. Second, akin to the opposite sign in the IC eq.~\eqref{Eq:B_TCA_OCT}, the MF starts with positive values, and then flips sign when $\delta v$ does, as was the case for ISO IC (this can be checked with the top plots of Figure~\ref{Fig:OCT_vs_ISO}). Third, the smaller the mode is, the more dominant the oscillatory part of ${\cal N}_2$ (and hence of $\Theta_2$) becomes due to diffusion damping, thus the more rapid the oscillations in $a^2{\cal B}$ are. This is evident from the orange ($k=1$ Mpc$^{-1}$) case, as well as from the $k=10$ Mpc$^{-1}$, which we omit from the plot to avoid obscuring the other curves. Finally, we have checked that increasing $r_v$ to $10^3$ increases $a^2{\cal B}$ slightly (${\cal B}_1 = 1.12\times10^{-26}$ G), and the resulting fit to the data is worsened. Thus, it is safe to conclude that the octupole IC is even less plausible to generate the required $B_{\text{seed}}$.

\begin{figure}[!htb]
	\centering
 	\includegraphics[width = 0.7\textwidth]{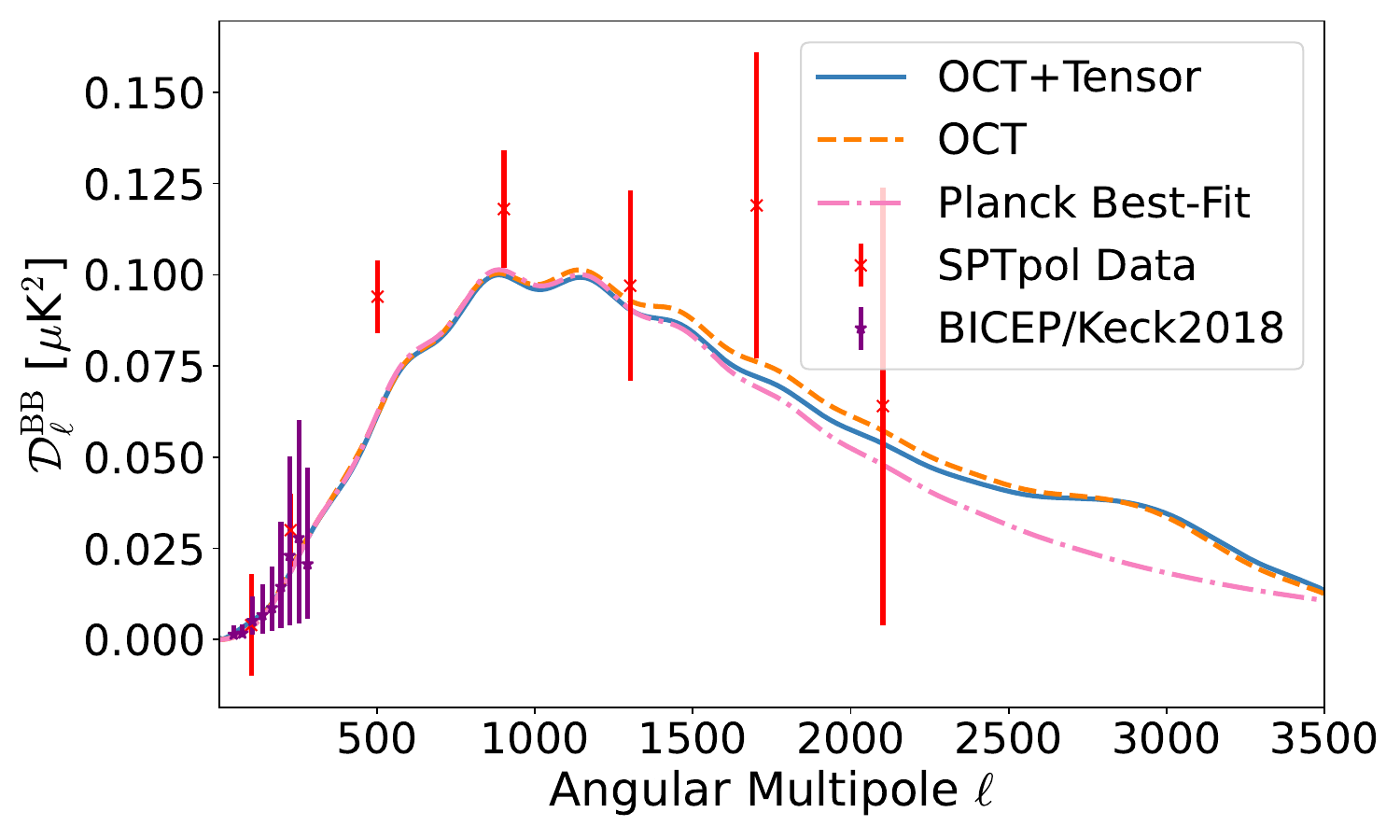}
 	\caption{\justifying Best-fit power spectrum of CMB polarization $B$-modes for the case of OCT ICs compared to data. The latter includes  data points from BICEP/Keck~\cite{BICEP_Keck} (purple star-shaped) and ones from SPTpol~\cite{SPTpol} (red x-shaped). The solid blue curve is the total lensed $BB$ spectrum coming from the presence of both $\mathcal{V}$-modes and Tensor modes. The dashed orange curve is for the total lensed spectrum due to the presences of $\mathcal{V}$-modes only. The pink dot-dash curve is the Planck $\Lambda$CDM best-fit lensed spectrum.}
 	\label{Fig:OCT_BB_lin_tens}
 \end{figure}

However, the resulting best-fit $BB$ spectrum is more compatible with data from BICEP/Keck and SPTpol than in the case of ISO IC. In fact, this holds even when Tensor modes are included, which gives the best-fit parameters listed in the 5$^{\text{th}}$ row of Table~\ref{Table: Best-Fits}. We show this result in Figure~\ref{Fig:OCT_BB_lin_tens}, where we plot the best-fit lensed $BB$ spectrum from the OCT IC alone (orange dash curve) and including tensor modes (blue solid curve) and compare it to the Planck $\Lambda$CDM best-fit spectrum (pink dot-dash curve) and data points from BICEP/Keck~\cite{BICEP_Keck} and SPTpol~\cite{SPTpol}. We should stress that this outcome does not mean a detection of primordial $\mathcal{V}$-modes, rather a hint that such an IC could be worth investigating with a more proper data analysis, which is beyond the scope of the current work. 

\textbf{Summary:} We investigated the possibility of having a non-zero IC on the neutrino octupole as a way to generate PMFs from $\mathcal{V}$-modes. Neutrino octupole sources the quadrupole, which also sources photon quadrupole through $\mathcal{V}$-modes. Then, the photon quadrupole creates a difference in its dipole w.r.t that of baryons, thus generating the electric field needed to source the MF. However, the resulting amplitude of MFs is too small to act as $B_{\text{seed}}$ needed to explain astrophysical MFs, making this mechanism less likely to be a magnetogenesis one. Nonetheless, the resulting best-fit CMB $BB$ spectrum has a similar compatibility with data to that of the Planck best-fit $\Lambda$CDM model. This motivates further investigation of such IC.  

\subsection{Sourced Mode Initial Condition}

In the previous two ICs, we have altered the neutrino sector in a non-trivial and unconventional way in an attempt to produce a $B_{\text{seed}}$ through the presence of $\mathcal{V}$-modes. Given that this was not achieved, we now attempt an even more unconventional and speculative IC, which we call the Sourced Mode (SMD) IC. We are considering such a scenario precisely to highlight the difficulty of producing PMFs from $\mathcal{V}$-modes, while the presence of the latter could still be compatible with data. 

In this IC, we seek to directly source the $\mathcal{V}$-modes with the anisotropic stress of some Dark species (DS). The latter are expected to be present in Super-symmetric extensions of the Standard Model of Particle Physics~\cite{SUSY1,SUSY2,SUSY3,SUSY4}, or in higher dimensional theories such as String Theory~\cite{String1,String2,String3,String4}. In principle, by directly sourcing the $\mathcal{V}$-modes, one should be able to produce a stronger velocity difference between photons and baryons, and thus a stronger $B_{\text{seed}}$. As a toy model for this scenario, let us assume that, on super-horizon scales, the anisotropic stress of this DS increases sharply at some conformal time $\eta_*$. In other words, the DS' anisotropic stress takes the form
\be
\pi_s= \pi_*\delta(\eta-\eta_*)
\label{Eq:DS_AnisotropicStress}
\ee
where $\pi_*$ is a constant (in units of Mpc) and $\delta(\eta-\eta_*)$ is a Dirac delta.

To understand the initial behaviour of all the relevant quantities, we first note that near the time $\eta_*$, the DS and neutrino velocities dominate the r.h.s of eq.~\eqref{Eq:Constraint_Eq} at leading order in TCA. This means that, for $\eta > \eta_*$, eq.~\eqref{Eq:Constraint_Eq} takes the form\footnote{In deriving this expression, we use the equivalent of eq.~\eqref{Eq:GenEuler} for DS after inserting eq.~\eqref{Eq:DS_AnisotropicStress}.}

\be
k^2\Phi = \frac{8\Omega_{\nu}}{\Omega_r\eta^2}\tilde{v}_{\nu}+k^2\Phi_*\left(\frac{\eta_*}{\eta}\right)^2,
\label{Eq:Constraint_SMD}
\ee
where $\Phi_*$ is the value of $\Phi$ just after $\eta_*$. Note that $\Phi_*$ is related to $\Phi_0$ of the previous ICs by
\be
\Phi_* = \Phi_0\left(\frac{a_{\text{ref}}}{a^2}\right),
\ee
where $a_{\text{ref}}$ is an arbitrary reference scale factor that we choose to be $a_{\text{ref}}=10^{-4}$. Next, we need the evolution equation for the neutrino velocity (i.e. dipole) on super-horizon scales ($k\eta\ll1$). Using eqs.~\eqref{Eq:Nu_dipole} and~\eqref{Eq:Nu_quad} (neglecting the octupole contribution) and inserting eq.~\eqref{Eq:Constraint_SMD}, we find

\be
\tilde{v}_{\nu}{}''+\frac{8\Omega_{\nu}}{5\Omega_r\eta^2}\tilde{v}_{\nu} = -\frac{k^2}{5}\Phi_*\left(\frac{\eta_*}{\eta}\right)^2.
\label{Eq:Evol_Nu_dipole_SMD}
\ee
The general solution of this differential equation can be found analytically to be

\be
\tilde{v}_{\nu} = a_{\nu}\left\{\sqrt{\frac{\eta}{\eta_*}}\left[\cos(x) - b_{\nu}\sin(x)\right]-1\right\},
\label{Eq:Sol_vnu_SMD}
\ee
where
\be
a_{\nu} = \frac{k^2\Phi_*\eta_*^2\Omega_r}{8\Omega_{\nu}}, \ b_{\nu} = \sqrt{\frac{5\Omega_r}{17\Omega_{\nu}+5\Omega_{\gamma}}}
\label{Eq:a_b}
\ee
and
\be
x = \frac{1}{2b_{\nu}}\ln{\frac{\eta}{\eta_*}}.
\label{Eq:x}
\ee
Inserting the solution~\eqref{Eq:Sol_vnu_SMD} into eq.~\eqref{Eq:Constraint_SMD}, we get for $\eta > \eta_*$,

\be
\Phi = \Phi_*\left(\frac{\eta}{\eta_*}\right)^{3/2}\left[\cos(x) - b_{\nu}\sin(x)\right].
\label{Eq:Sol_Phi_SMD}
\ee

The above presented solutions for $\tilde{v}_{\nu}$ and $\Phi$ are at leading order in TCA. Given that the photo-baryonic contribution is negligible at the epoch considered here, we need to go to NLO in TCA to find the expressions for photon and baryon velocities, and then eventually for $a^2{\cal B}$. Applying approximations eqs.~\eqref{Eq:NLO_approx} and~\eqref{Eq:deltav_vf_OCT}, we find
\be
-\frac{\tau{}'\delta v}{R} = -\frac{4k^2}{15}\left[\eta^{-1}\int d\eta\Phi\tau{}'^{-1}+\Phi\tau{}'^{-1}\right]\,.
\label{Eq:delav_SMD}
\ee
Finally, after solving eq.~\eqref{Eq:delav_SMD} and inserting it in eq.~\eqref{Eq:B_final}, we get
\be
a^2{\cal B} = a_{{\cal B}}k^3\eta^{5/2}\left[b_{{\cal B}}\cos(x)-c_{{\cal B}}\sin(x)\right],
\label{Eq:Sol_B_SMD}
\ee
where
\be
a_{{\cal B}} = \frac{32\pi Gm_p^2\Phi_*{\cal H}_0 \Omega_r^{2}\left(65\Omega_{r}+6\Omega_{\nu}\right)^{-1}}{45e\Omega_b\sigma_T\left(25\Omega_{r}-16\Omega_{\nu}\right)}\eta_*^{3/2},
\label{Eq:a_B}
\ee
\be
b_{{\cal B}} = 20\sqrt{\Omega_{r}}\left(65\Omega_r-16\Omega_{\nu}\right),
\ee
and
\be
c_{{\cal B}} = \sqrt{\frac{20}{12\Omega_{\nu}+5\Omega_r}}\left(325\Omega_r^2-30\Omega_r\Omega_{\nu}+96\Omega_{\nu}^2\right).
\ee

\begin{figure}[!htb]
		\begin{subfigure}[t]{0.53\textwidth}
			\hspace{-1.9cm}
			\centering
			\includegraphics[width=\textwidth]{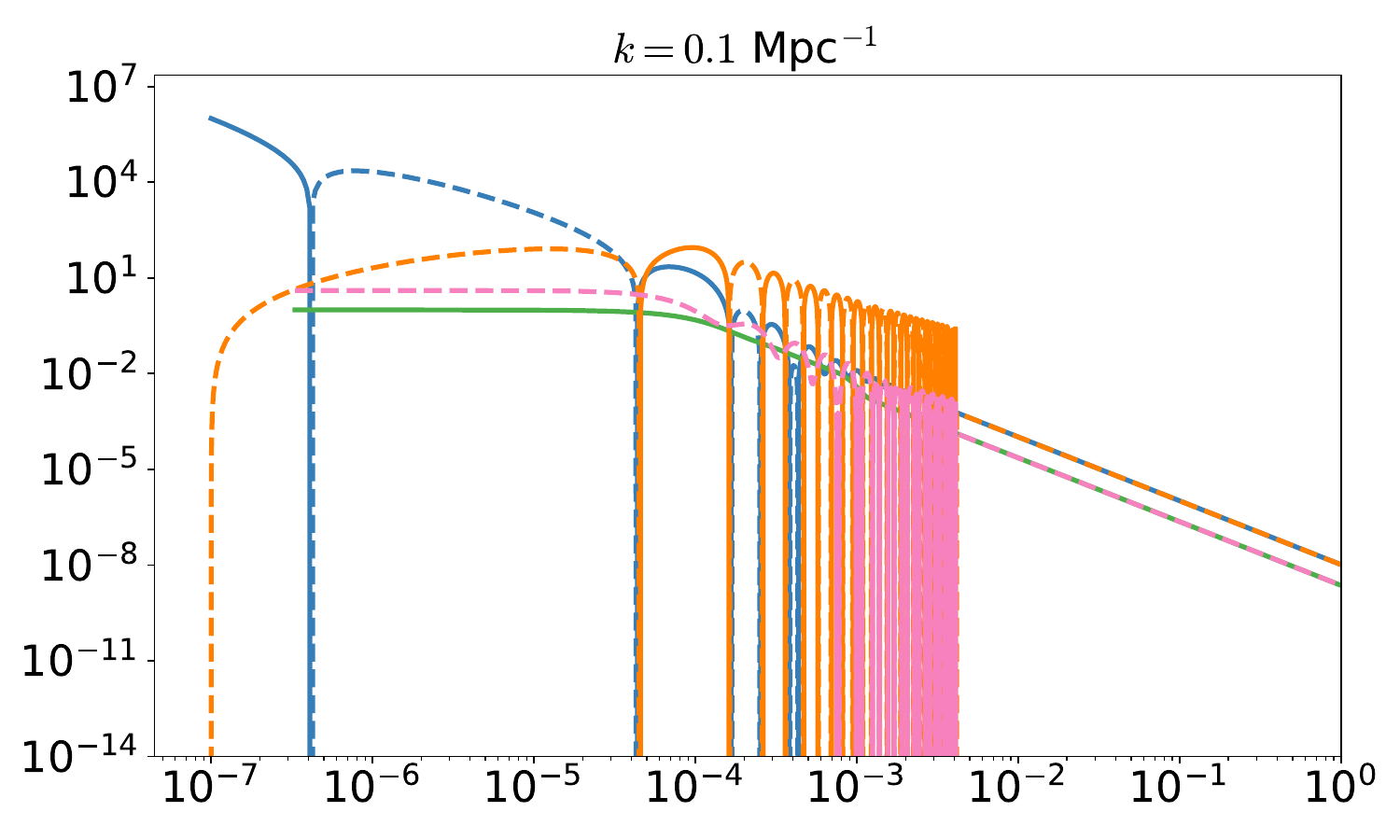}
		\end{subfigure}%
		\hfill
		\begin{subfigure}[t]{0.53\textwidth}
			\hspace{-1.9cm}
			\centering
			\includegraphics[width=\textwidth]{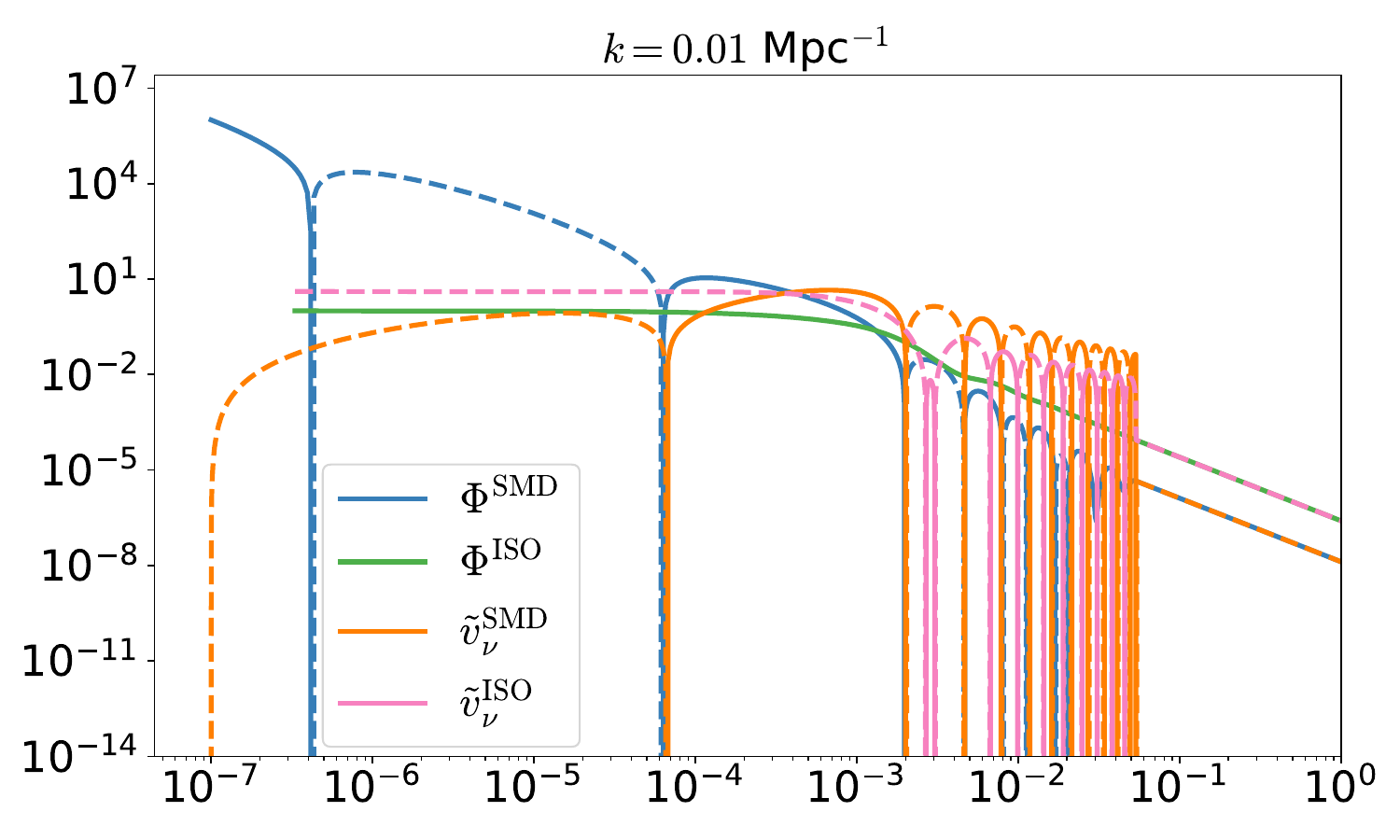}
		\end{subfigure}
		\vspace{0.5cm}
		\begin{subfigure}[t]{0.53\textwidth}
			\hspace{-1.9cm}
			\centering
			\includegraphics[width=\textwidth]{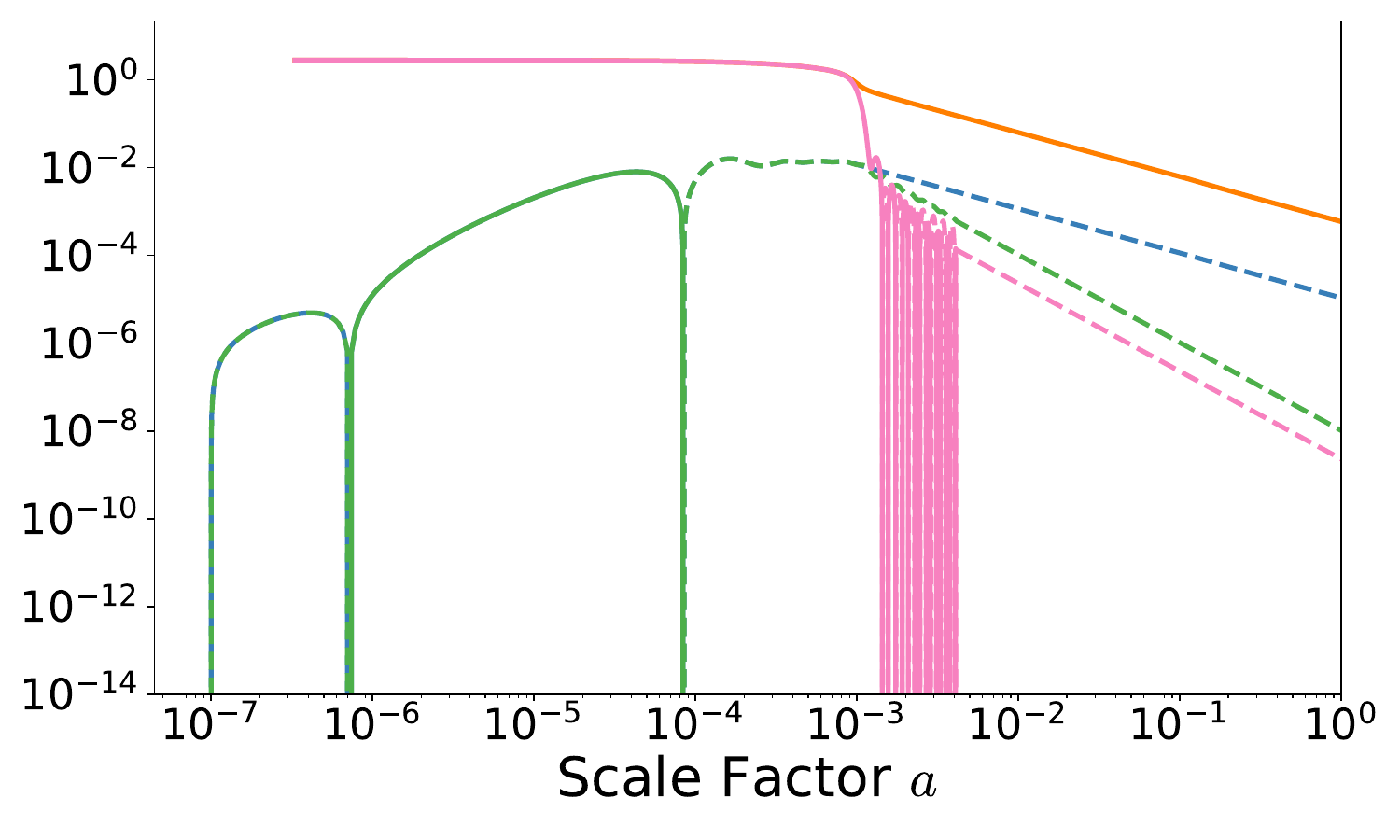}
		\end{subfigure}%
		\hfill
		\begin{subfigure}[t]{0.53\textwidth}
			\hspace{-1.9cm}
			\centering
			\includegraphics[width=\textwidth]{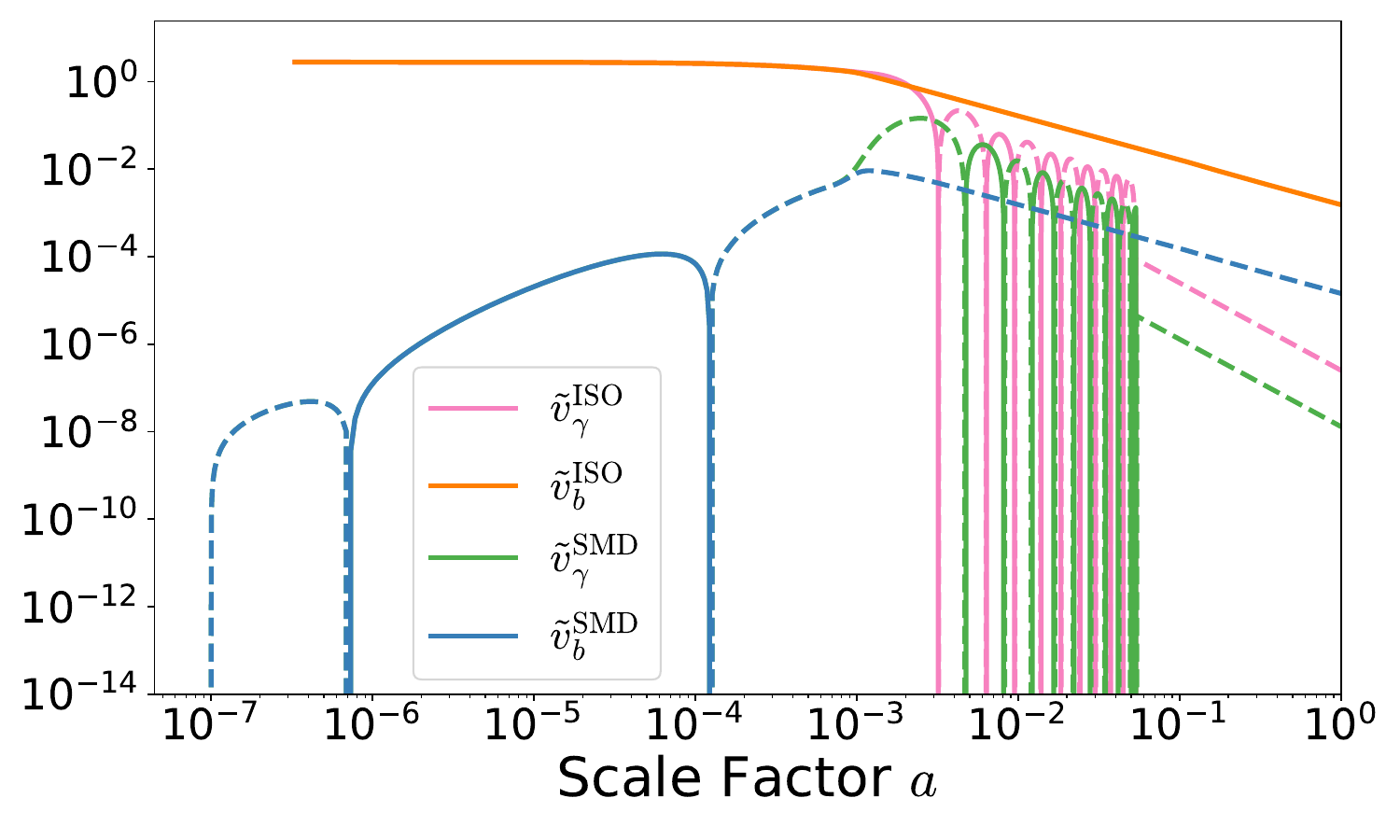}
		\end{subfigure}
		\caption{\justifying Evolution of $\Phi,\ \tilde{v}_b,\ \tilde{v}_{\nu},$ and $ \tilde{v}_{\gamma}$ (relative to $\Phi_0$) with the scale factor $a$ for SMD IC. Each term is plotted in comparison with the corresponding term for ISO IC. The plots on the left are for $k = 0.1$ Mpc$^{-1}$, while the ones on the right are for $k = 0.01$ Mpc$^{-1}$. The dashed lines correspond to negative values, while the solid lines to positive ones. In producing these plots, we set $r_v=10^{3}$, $n_v=0$ and the Planck best-fit for the $\Lambda$CDM cosmological parameters. We also set $\eta_*$ to correspond to a redshift $z_*=10^{7}$. The RSA is applied when oscillations become large at late times.}
\label{Fig:SMD_vs_ISO}
	\end{figure}

The evolution of the relevant quantities with $a$ is presented in Figure~\ref{Fig:SMD_vs_ISO}. We choose $\eta_*$ to correspond to a redshift $z_*=10^{7}$, so that it's early enough for TCA to apply, but still after neutrino decoupling. The first thing to notice is the oscillatory behavior at early times, as expected from the solutions eqs.~\eqref{Eq:Sol_vnu_SMD} and~\eqref{Eq:Sol_Phi_SMD}, with more extended oscillations for larger scales ($k=0.01$ Mpc$^{-1}$). By comparing the top to bottom plots, we can see that the amplitude of the photo-baryonic fluid velocity near $\eta_*$ is negligible compared to that of neutrinos. This justifies the assumption made before in deriving the ICs. 

When the small scale mode ($k=0.1$ Mpc$^{-1}$) enters the horizon, the photo-baryonic fluid's velocity stabilizes until photon diffusion becomes dominant. Then, the photon velocity starts decaying, but without the oscillations apparent in both OCT and ISO ICs. This is associated with the slow decay of $\Phi$, which can be seen in the top left plot of Figure~\ref{Fig:SMD_vs_ISO}. However, for the large scale mode, the photon and baryon velocities follow a similar evolution to that of the OCT case after horizon crossing. This occurs because for large scale modes $\Phi$ has decayed enough to make its effect less impactful, and since there's no ISO condition, the system will be similar to the OCT case.
\subsubsection{Best-Fit MF and CMB BB Spectrum}

\begin{figure}[!htb]
	\centering
 	\includegraphics[width = 0.7\textwidth]{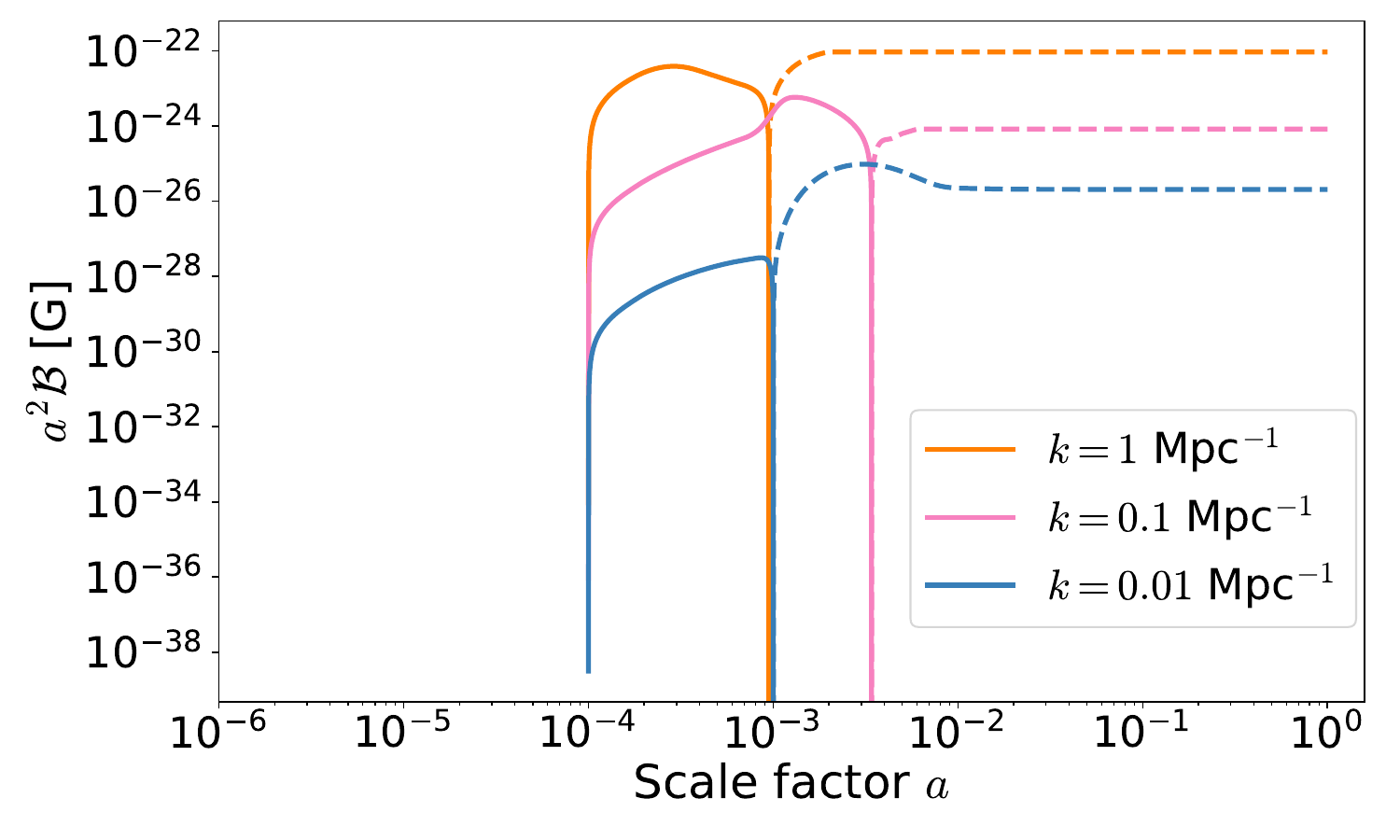}
 	\caption{\justifying Same as Figure~\ref{Fig:ISO_B_vs_a} for SMD IC, excluding the $k=10$ Mpc$^{-1}$ mode.}
 	\label{Fig:SMD_B_vs_a}
 \end{figure}

When first looking for the best-fitting parameters of this ICs to the Planck, SPT and BAO data, we set $z_*=10^7$. However, we found the resulting MF to be much smaller than the OCT case, and thus we decreased $z_*$ to $10^4$. This is close to the maximum redshift where the above mentioned approximations are still valid. We show the evolution of ${\cal B}$ in Figure~\ref{Fig:SMD_B_vs_a}.

We can see from that figure the sharp increase in ${\cal B}$ when $\Phi$ gets sourced by the DS. Then, the amplitude for the large scale mode increases almost linearly until it switches sign when the photon velocity becomes larger than the baryon one. On the other hand, the amplitude for $k=0.1$ Mpc$^{-1}$ continues to grow in accordance with the velocity difference between photons and baryons. However, close to $a= 4\times10^{-3}$, the increase in photon velocity (seen in Figure~\ref{Fig:SMD_vs_ISO}) causes a flip in the sign of ${\cal B}$, resulting in its stabilization at a negative value close to the end of photon decoupling.

\begin{figure}[!htb]
 	\centering
 	\includegraphics[width = 0.7\textwidth]{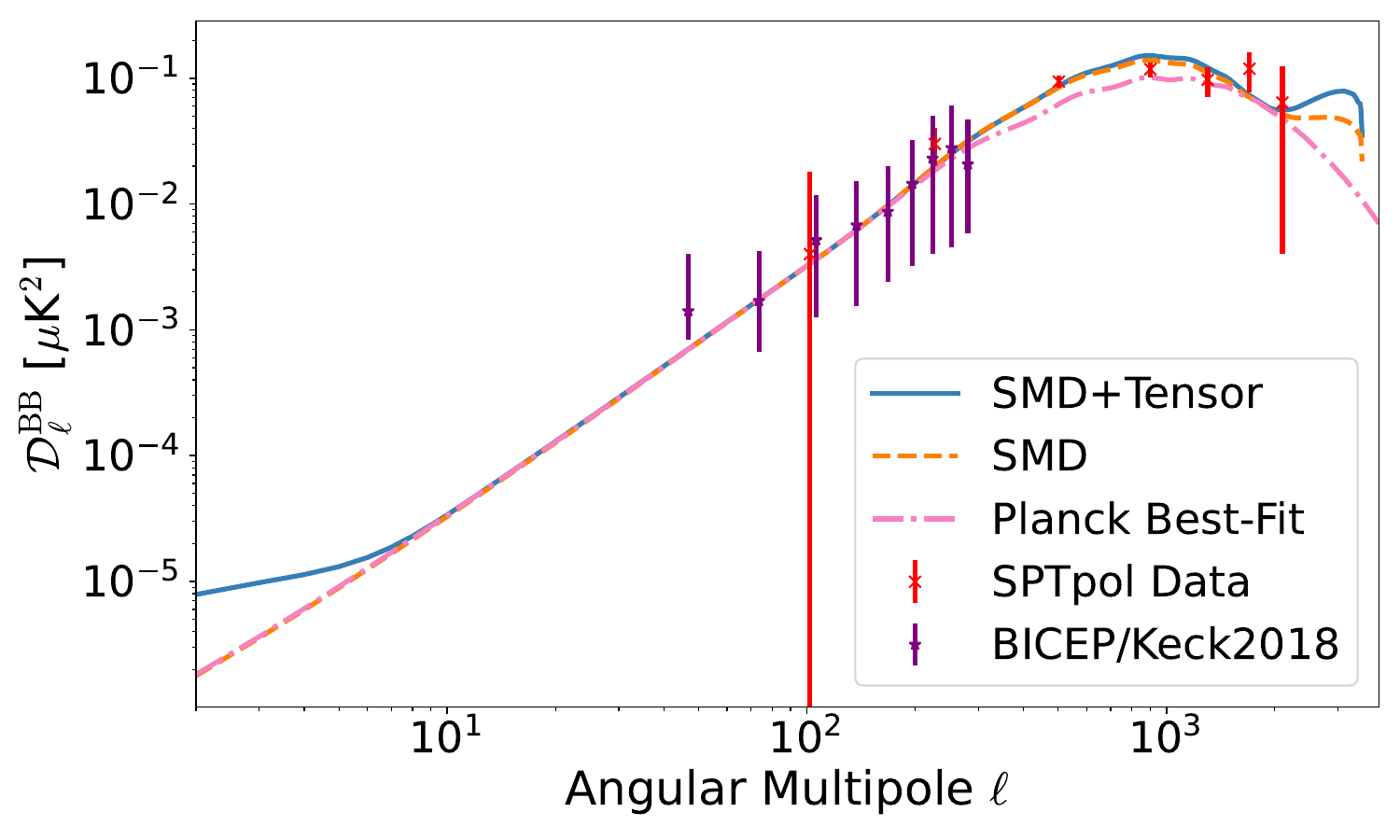}
 	\caption{\justifying Same as Figure~\ref{Fig:OCT_BB_lin_tens} but for the SMD IC and in logarithmic scale.}
 	\label{Fig:SMD_BB_loglog_tens}
 \end{figure}

More importantly for the current work, this IC produced ${\cal B}_1\sim{\cal O}(10^{-26})$G (see Table~\ref{Table: Best-Fits}), and thus is not a suitable magnetogenesis candidate. Nevertheless, unlike the ISO case, we see from Figure~\ref{Fig:SMD_BB_loglog_tens} that the resultant $BB$ spectrum (dashed orange) has a similar fitting power to the one of Planck best-fit $\Lambda$CDM case (dot-dash pink), even when including tensor modes (solid blue). Moreover, one can see interesting differences at large and small scales due to the presence of tensor and vector modes, respectively. As was the case of the OCT IC, a more detailed data analysis is needed in order to make a better judgment about this IC. 

\textbf{Summary:}
The presence of Dark Species in the early Universe could be a source of $\mathcal{V}$-modes. Using a toy model where the anisotropic stress of these species increases abruptly at a given redshift, we study the resultant evolution of $\mathcal{V}$-modes, and their direct effect on the velocity of photons and baryons. The latter two develop velocity differences that create a non-zero MF. However, as was the case in the previous ICs, the amplitude of this MF is not large enough to be a candidate for $B_{\text{seed}}$. Nevertheless, the CMB polarization $B$-mode spectrum shows similar behavior to the Planck best-fit $\Lambda$CDM case when compared to data from BICEP/Keck and SPTpol.
\begin{table}[h]
			\centering
	\begin{tabular}{c|c|c|c|c}

		  & $r_v$ & $\ n_v \ $ & $10^2r$ & $10^{26}{\cal B}_1$[G]\\
		  \hline
		  ISO & 0.001 & 0.4 & --- & 0.5\\
		  
		  OCT & 15.3 & 6.7 & ---& 0.2\\
		  
		  SMD & 6.7 & 1.2 & ---& 2.6\\
		  
		  ISO + Tensor & 0.003 & 0.4 & $2.5$ & 0.8 \\
		  OCT + Tensor & 2.1 & 8.0 & $2.6$ & 0.2\\
		  SMD + Tensor & 7.8 & 1.5 & $0.04$& 4.5\\
		  \hline
	\end{tabular}
\caption{\justifying Best-fit values for the amplitude of $\mathcal{V}$-modes' power spectrum ($r_v$), its spectral index ($n_v$), the amplitude of tensor perturbations' power spectrum ($r$), when applicable, and the amplitude of the MF (as given by eq.\eqref{Eq:Smoothed_B}) for the different ICs considered in this work. The first three rows are for only $\mathcal{V}$-modes, while the last three ones are for including gravitational tensor modes on top of them.}
\label{Table: Best-Fits}
\end{table}

\section{Comparison of CMB Spectra}
\label{Sec:CMB_Spectra}
Having seen the detailed mechanism of each IC, and how they can produce PMFs, we now turn briefly to the behavior of the CMB spectra in each IC. From eq.~\eqref{Eq:CMB_Spect}, we see that the presence of $\mathcal{V}$-modes sources the CMB anisotropies. There will be a similar expression coming from scalar and tensor modes, such that the final CMB spectrum is the sum of all three:
\be
C_{\ell}^\mathrm{XY,tot} = C_{\ell}^\mathrm{XY,s} + C_{\ell}^\mathrm{XY,v} + C_{\ell}^\mathrm{XY,t}
\label{Eq:CMB_tot}
\ee
where $X,Y=\{T,E,B\}$ and s, v and t stand for scalar, vector and tensor modes, respectively. We provide a qualitative description of these spectra without detailing the quantitative one.

\begin{figure}[!htb]
		\begin{subfigure}[t]{0.53\textwidth}
			\hspace{-1.9cm}
			\centering
			\includegraphics[width=\textwidth]{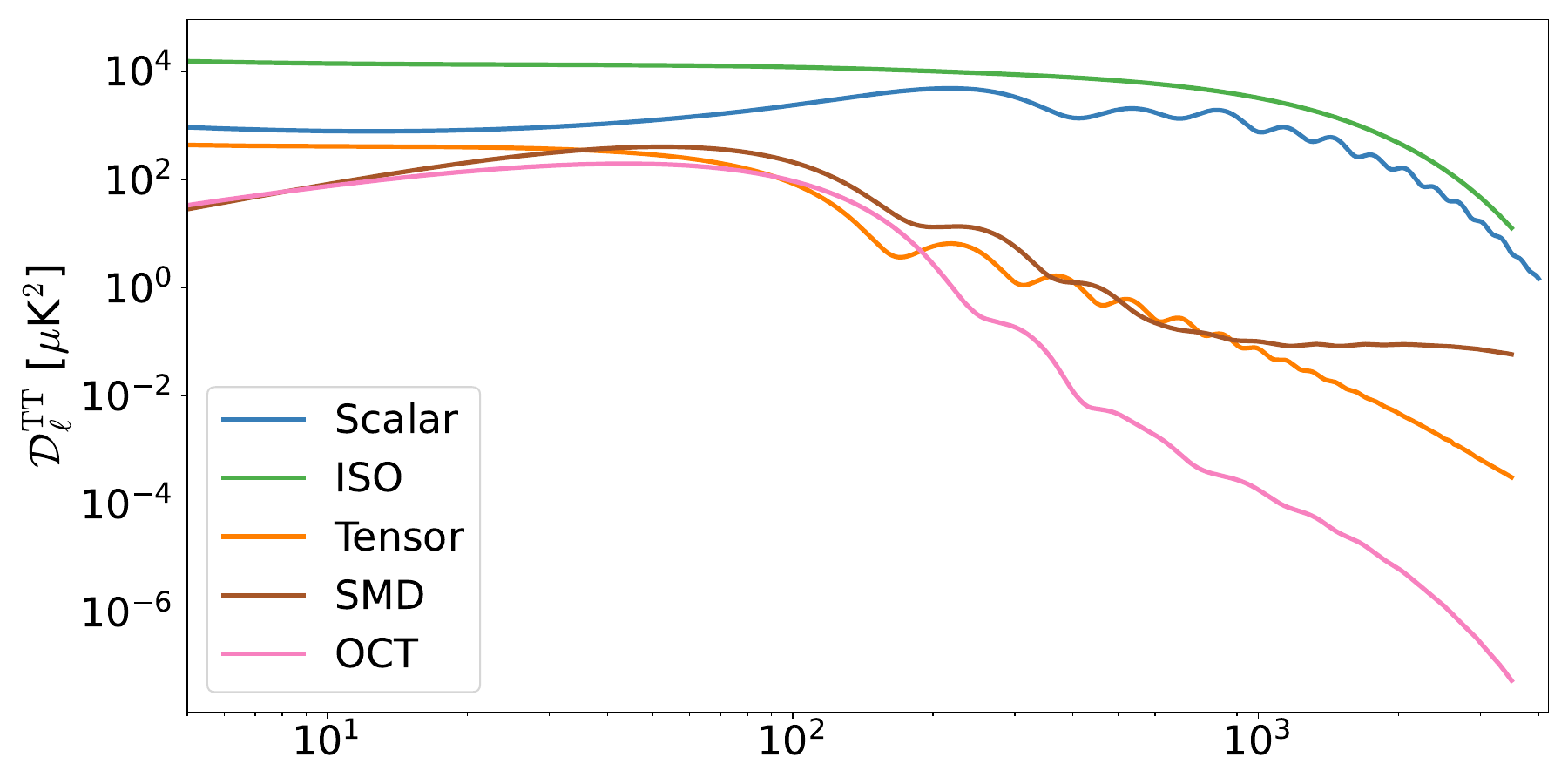}
		\end{subfigure}%
		\hfill
		\begin{subfigure}[t]{0.53\textwidth}
			\hspace{-1.9cm}
			\centering
			\includegraphics[width=\textwidth]{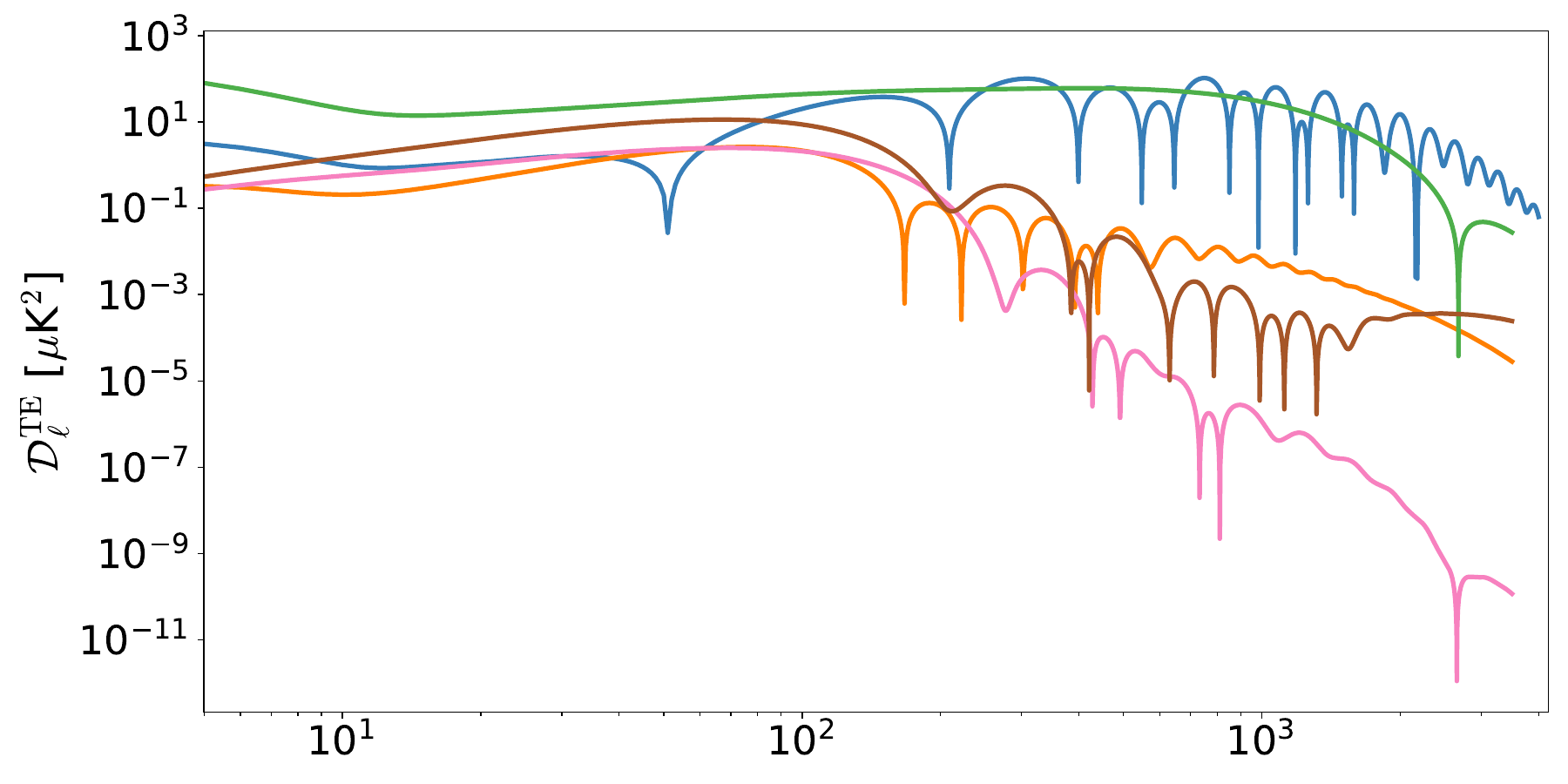}
		\end{subfigure}
		\vspace{0.5cm}
		\begin{subfigure}[t]{0.53\textwidth}
			\hspace{-1.9cm}
			\centering
			\includegraphics[width=\textwidth]{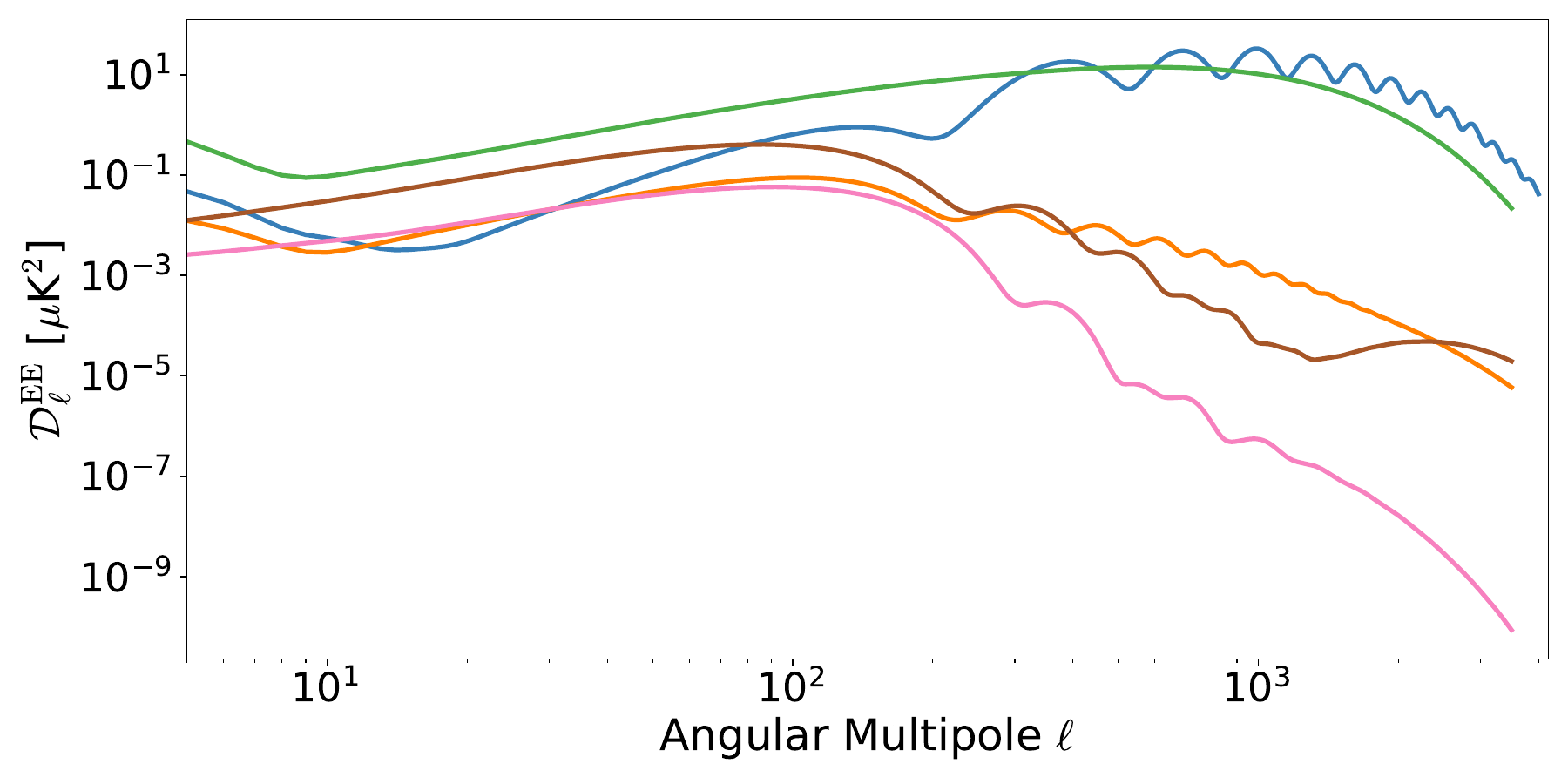}
		\end{subfigure}%
		\hfill
		\begin{subfigure}[t]{0.53\textwidth}
			\hspace{-1.9cm}
			\centering
			\includegraphics[width=\textwidth]{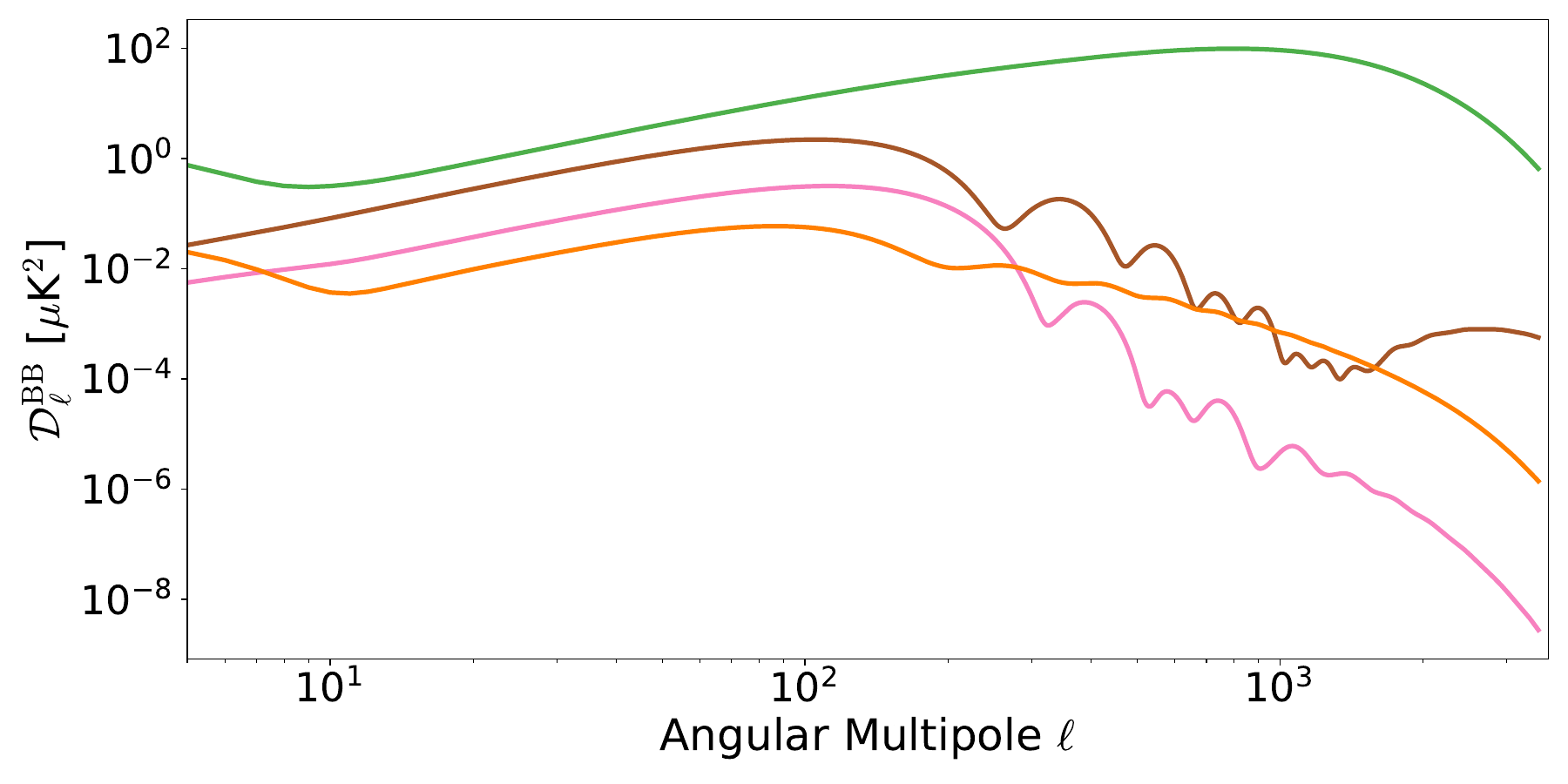}
		\end{subfigure}
		\caption{\justifying CMB spectra, ${\cal D}_{\ell}$, for the three different $\mathcal{V}$-modes considered in this work compared to those of Scalar and Tensor modes. For illustration, we set the $\Lambda$CDM cosmological parameters to those given by the Planck best-fit~\cite{Planck_Primary},
        $r_v=r=1$ and $n_v=n_t=n_s$ where $n_t$ is the tensorial spectral index. For the SMD case, we set $z_*=10^7$.}
\label{Fig:CMB_Spectra}
\end{figure}

We show the behavior of $TT,\ TE,\ EE$ and $BB$ spectra, as a function of multipoles $\ell$, for each IC in Figure~\ref{Fig:CMB_Spectra}, expressed in terms of
\be
{\cal D}_{\ell}^{XY} = \frac{\ell(\ell+1)}{2\pi}C_{\ell}^{XY}.
\label{Eq:D_ell}
\ee
We also show, for each CMB spectrum, the behavior of the latter due to scalar (blue) and tensor (orange) modes. In order to have the s, v and t-modes on equal footings, we set $r_v=r=1$ and $n_v=n_t=n_s$, so that the three have the same amplitude and spectral indices, respectively. Even with this setting, the three modes produce different signatures on the CMB. This is the result of not only the geometrical difference between the three types of modes, but for the $\mathcal{V}$-modes case also the way in which each IC has been sourced.

Focusing first on the ISO case, we can see that in all four CMB spectra, this IC features barely any acoustic peaks, unlike the OCT and SMD cases. Moreover, the ISO case shows comparable amplitude to the s-mode in $TT,\ TE$ and $EE$, explaining why when constrained with data, this IC had the smallest $r_v$. This behavior is due to the ISO case being the one which directly affects the CMB dipole the most compared to the other. Furthermore, the sharp rise in the $BB$ spectrum at intermediate to large $\ell$s is special to this IC, showing that data from that $\ell$ range will be crucial in putting upper limits on its amplitude.

On the other hand, the OCT and SMD cases show clear acoustic peaks starting from $\ell\gtrsim200$. Another similarity between them is the comparable behavior and amplitude, specially in $TT$, at large scales ($\ell\lesssim200$). This is related to what was described in Figure~\ref{Fig:SMD_vs_ISO}, where we saw the similarity between SMD and OCT ICs for large scale modes. However, one apparent difference between the two ICs is that the SMD results in more strongly decaying spectra compared to the OCT case. Finally, and most importantly for experiments measuring $BB$ spectra, both ICs show comparable, even slightly larger, amplitudes compared to the t-modes spectra at large and intermediate scales $\ell\lesssim1000$. This means that the slightest presence of $\mathcal{V}$-modes can bias the conclusions made about primordial t-modes, and therefore about any inflationary mechanism~\cite{Inflation1,Inflation2,Inflation3}.

\section{Conclusion}
\label{Sec:Conclusion}
In this work, we revisited the possibility of gravitational $\mathcal{V}$-modes being present in the early Universe, and how they can generate PMFs. The main mechanism is for the $\mathcal{V}$-modes to produce a difference in velocities between baryons and photons. This then creates an electric field which in turn produces the PMF.  We considered three different ICs that could source such $\mathcal{V}$-modes long enough to, in principle, produce PMFs.

The first IC, the isocurvature, is the generation of $\mathcal{V}$-modes due to a velocity difference between photon and neutrinos after neutrino decoupling. Such a difference alters the evolution of both species' anisotropic stress in a way that can source $\mathcal{V}$-modes. This IC has been considered in previous works~\cite{Lewis:2004kg,Itchiki_V-modes,Paoletti_theory,Rebhan:1991sr,Rebhan:1994zw}, which we managed to reproduce. Specifically, we highlighted interesting features in the temporal evolution of the different species involved (Figure~\ref{Fig:ISO_Perts_vs_eta}), and showed how that can explain the evolution of MFs (Figure~\ref{Fig:ISO_B_vs_a}). The result is that such IC cannot produce large enough seed MF to explain the origin of all the observed  astrophysical MFs. The later needs a $B_{\text{seed}}\sim10^{-14}$ G on scales of 1 Mpc, whereas the ISO IC produced an amplitude almost 12 orders of magnitude smaller (Table~\ref{Table: Best-Fits}). We also showed that increasing the amplitude of $\mathcal{V}$-modes further to increase that of PMFs is not allowed. This is because the resultant $BB$ spectrum of this IC from best-fitting its parameters to data from Planck~\cite{Planck_Lensing,Planck_Primary}, SPT-3G 2018~\cite{Balkenhol_etal} and BAO~\cite{BOSS_DR7,BOSS_DR12,eBOSS_DR16} is already in stark violation of data from SPTpol~\cite{SPTpol} (Figure~\ref{Fig:ISO_BB_lin}).

The second IC we considered is called the neutrino octupole. As the name suggests, we assumed the presence of a non-vanishing neutrino octupole in the early Universe, which then acts as a source for the quadrupole, i.e. the anisotropic stress, thus sourcing $\mathcal{V}$-modes. Following the same analysis as for the previous IC, we compared the behavior of all relevant quantities between the two ICs (Figure~\ref{Fig:OCT_vs_ISO}) and the resultant MF's evolution (Figure~\ref{Fig:OCT_B_vs_a}). Again, the resultant $B_{\text{seed}}$ in this case is ${\cal O}(10^{-27})$ G, too small to account for all MFs observed, making this IC also less likely to be a magnetogenesis mechanism. However, the resultant best-fit $BB$ spectrum shows similar fitting power as the $\Lambda$CDM Planck best-fit one. This suggests that the presence of such $\mathcal{V}$-modes might not be entirely excluded, and a further data analysis of this IC is needed in order to put constraints on its parameters.

The last IC studied is named the Sourced Mode IC. In here, we assumed the presence of Dark species in the early Universe with an anisotropic stress that abruptly sources $\mathcal{V}$-modes at a particular redshift $z_*$. After comparing the temporal behavior of the quantities involved to the previous ICs (Figure~\ref{Fig:SMD_vs_ISO}) and looking at the evolution of the MF (Figure~\ref{Fig:SMD_B_vs_a}), we also concluded that this IC cannot be a good candidate for magnetogenesis in the early Universe (the MF amplitude in this case is $\sim{\cal O}(10^{-26})$ G). Nonetheless, as in the OCT case, we found that this IC can produce a $BB$ spectrum that is not in obvious violation of data from BICEP/Keck~\cite{BICEP_Keck} or SPTpol~\cite{SPTpol}. Therefore, a more thorough data analysis is needed to better constrain this IC.

We investigated such speculative ICs precisely to highlight the difficulty of generating seed MFs from $\mathcal{V}$-modes. Therefore, another mechanism would still be needed for that purpose. However, we also highlight that the smallest presence of $\mathcal{V}$-modes can produce $BB$ spectra with similar power to primordial tensor modes. Hence, it is crucial to put upper limits on the amplitude of $\mathcal{V}$-modes in order to make sure that our conclusions from $BB$ measurements on inflationary mechanisms is not biased. Moreover, $\mathcal{V}$-modes are an intrinsic property of Riemannian geometry, the fundamental framework of General Relativity (and most gravitational theories). If their primordial amplitude is exactly zero, this is a fine-tuning problem that needs to be addressed, as was the case for the flatness problem in inflation~\cite{Inflation1,Inflation2,Inflation3}. Otherwise, this amplitude should be constrained with data in order to make sure that a detected primordial $BB$ spectrum is not confused with other mechanisms.
\section*{Acknowledgments}

We thank Silvia Galli from IAP for fruitful discussions during all stages of this project. This project has received funding from the European Research Council (ERC) under the European Union’s Horizon 2020
research and innovation program (grant agreement No 101001897). This work has also made use of the Infinity Cluster hosted by IAP.

\appendix
\section{Vector Perturbations Harmonics in Flat Space}
In this appendix, we  briefly present technical details related to the content of Section~\ref{Sec:Norm_Modes_Exp}. These details summarize some of the expressions collected in ref.~\cite{PitrouModes}.

First, the propagation direction convention used in~\cite{PitrouModes} is
\be
\bar{\boldsymbol{n}}\equiv-\boldsymbol{n}
\ee
where $\boldsymbol{n}$ is the observed direction. This, together with the sign convention in eq.~\eqref{Eq:Q^pm}, ensures that the physically relevant eqs.~\eqref{Qini} and~\eqref{Qijnij} are the same as in the convention of~\cite{HuAndWhite}.

Second, the radial functions ${}_s\bar{\alpha}_{\ell}^{(\ell'm)}$ appearing in eq.~\eqref{DefGslm} are related to the ones of the observed direction convention by
\be
{}_s\bar{\alpha}_{\ell}^{(\ell'm)}(x) = {}_s\alpha_{\ell}^{\ell'm}(-x),
\label{Eq:s_alpha^lm_l'1}
\ee
and in a flat background they take the form
\bea
{}_s\alpha_{\ell}^{\ell'm}(x) \equiv  &\sum_L\sqrt{\frac{4\pi(2L+1)}{(2\ell+1)(2j+1)}}
&\times{}^sC_{\ell Lj}^{m0m}j_L(x)(i)^{L+j-\ell},
\label{Eq:s_alpha^lm_l'2}
\eea
where $j_L$ are the spherical Bessel functions of the first kind, and
\be
{}^sC_{\ell_1\ell_2\ell_3}^{m_1m_2m_3}\equiv (-1)^{m_1+s}\sqrt{\frac{(2\ell_1+1)(2\ell_2+1)(2\ell_3+1)}{4\pi}}
\begin{pmatrix}
\ell_1 & \ell_2 & \ell_3 \\
s & 0 & -s
\end{pmatrix}
\begin{pmatrix}
\ell_1 & \ell_2 & \ell_3 \\
-m_1 & m_2 & -m_3
\end{pmatrix},
\label{Eq:Wigner3j}
\ee
with the 2$\times$3 matrices being the Wigner 3-j symbol. 

Third, the general relationship between normal modes and eq.~\eqref{DefGslm} for tensors of any rank is (suppressing the dependence on ${\bm k}$ for simplicity)
\be
\bar{Q}_{I_j}^{(jm)}(-\chi \bar{\boldsymbol{n}})\equiv\sum_{s=-j}^{j}{}_sg^{(jm)}{}_s\bar{G}^{(jm)}(\chi,\bar{\boldsymbol{n}})\hat{n}_{I_j}^s(\bar{\boldsymbol{n}}),
\label{Eq:Q_G_relation}
\ee
where $I_j=i_1...i_j$. The numerical factors ${}_sg^{(jm)}$ take the following values for the cases most relevant for this work:
\be
{}_0g^{(1m)}=\mp{}_{\pm1}g^{(1m)}=1\,,\ 
{}_0g^{(2m)}=\mp{}_{\pm1}g^{(2m)}=\frac{\sqrt{3}}{2}\,, \ 
{}_{\pm2}g^{(2m)}=\frac{1}{\sqrt{2}}\,.
\label{Eq:Num_Facs_Q_G}
\ee
In addition, $\hat{n}_{I_j}^s(\boldsymbol{n})$ constitutes a basis for $j$-rank tensors of spin $s$ fields, satisfying the orthonormality relation
\be
\hat{n}_{I_j}^{s}\hat{n}^{I_j}_{s'}=\delta_{ss'}d_{js}\,,
\label{Eq:Norm_hat_n}
\ee
where
\be
d_{js}=\frac{(j+s)!(j-s)!}{2^{|s|}(2j-1)!!j!}\,.
\label{Eq:Norm_hat_n_cont}
\ee
More specific to this work, these basis vectors take the following form
\be
\hat{n}_i=n_i; \quad \hat{n}_i^{\pm1} = \gr{e}_i^{\pm}
\label{Eq:hatn_i}
\ee
for $j=1$, and
\be
\hat{n}_{ij} = n_in_j-\frac{1}{3}g_{ij}\,; \ 
\hat{n}_{ij}^{\pm1} = \frac{1}{2}\left(\gr{e}_i^{\pm}n_j+\gr{e}_j^{\pm}n_i\right)\,; \ 
\hat{n}_{ij}^{\pm2} = \gr{e}_i^{\pm}\gr{e}_j^{\pm}\,,
\label{Eq:hat_n_ij_pm}
\ee
for $j=2$, where $g_{ij}$ is the spatial metric and $ \gr{e}^{\pm}$ are given by eq.~\eqref{Eq:Basis_vec}. By combining these definitions, one can derive eqs.~\eqref{Qini} and~\eqref{Qijnij}.

Finally, using the orthonormality relations,
\be
\int d^2\gr{n}\ {n}^i {n}_j = \frac{4\pi}{3}\delta^i_j, \ 
\int d^2\gr{n} \ \hat{n}^{i_1i_2}\hat{n}_{j_1j_2} = \frac{4\pi}{15}
\times\left[\delta^{i_1}_{j_1}\delta^{i_2}_{j_2} + \delta^{i_1}_{j_2}\delta^{i_2}_{j_1} - \frac{2}{3}g^{i_1i_2}g_{j_1j_2}\right],
\label{Eq:ortonormality_n_ij}
\ee
one can derive the relation between the anisotropic stress and the photon quadrupole~\eqref{Eq:Pim_def}.

\bibliography{Biblio.bib}

\end{document}